\documentclass[english,american,prx, twocolumn, longbibliography, superscriptaddress]{revtex4-2}
\usepackage[T1]{fontenc}
\usepackage[latin9]{inputenc}
\usepackage{amsmath}
\usepackage{amssymb}
\usepackage{graphicx}
\makeatletter

\providecommand{\tabularnewline}{\\}

\makeatother

\usepackage{babel}
\begin{document}
\title{Overdamped Lattice Dynamics of Sedimenting Active Cosserat Crystals}
\author{A. Bolitho}
\email{ab2075@cam.ac.uk }

\affiliation{DAMTP, Centre for Mathematical Sciences, University of Cambridge,
Cambridge CB3 0WA, United Kingdom}
\author{R. Adhikari}
\email{ra413@cam.ac.uk}

\affiliation{DAMTP, Centre for Mathematical Sciences, University of Cambridge,
Cambridge CB3 0WA, United Kingdom}
\begin{abstract}
Micropolar active matter requires for its kinematic description both
positional and orientational degrees of freedom. Activity generates
dynamic coupling between these kinematic variables that are absent
in micropolar passive matter, such as the oriented crystals first
studied by the Cosserat brothers. Here we study the effect of uniaxial
activity on the dynamics of an initially crystalline state of spheroidal
colloids sedimenting slowly in a viscous fluid remote from confining
boundaries. Despite  frictional overdamping by the fluid, the crystalline
lattice admits traveling waves of position and orientation. At long
wavelengths these obey a vector wave equation with Lamé constants
determined by the activity. We find that at least one polarization
mode of these waves is always unstable, leading to the melting of
the crystal. These results are elucidated by identifying an odd-dimensional
Poisson structure consisting of a Hamiltonian and an associated Casimir
invariant, where linear combinations of position and orientation are
identified as conjugate variables. Our results suggest that Poisson
structures may exist generally for active particles in slow viscous
flow and thereby allow equilibrium arguments to be applied in the
presence of these dissipative systems.
\end{abstract}
\maketitle

\section{Introduction}

The continuum mechanics of crystals comprised of oriented particles
was first studied theoretically by the Cosserat brothers in their
monograph of 1909 \citep{cosserat1909theorie}, \textit{``Theorie
des corps déformables''}. After an initial period of neglect, this
engendered the thriving field of micropolar and micromorphic continuum
mechanics, where the elementary material constituent is conceived
of having both a position and an orientation \citep{kafadar1971micropolar,eringen1967linear,altenbach2011mechanics}.
The requirement of invariance of the power expended under a rigid
motion necessarily couples the position and orientation degrees of
freedom and leads to a variety of macroscopic effects that have been
confirmed experimentally \citep{kroner1968mechanics,germain2020method,germain1973method,altenbach2011mechanics}. 

Active matter \citep{marchetti2013}, exemplified by suspensions of
active particles in a viscous fluid \citep{ebbens2010pursuit}, provides
a novel example of a mechanical system where both position and orientation
are relevant variables due to the breaking of microscopic rotation
symmetry \citep{lighthill1952squirming,brenner1963}. In contrast
to the Cosserat solid, however, active suspensions are comprised of
two components: the particles and the solvent. In the limit where
the hydrodynamics of the solvent are well modeled by the Stokes equations
of slow viscous flow, the particle motion is frictionally overdamped
and particle inertia plays no role in dynamical evolution. This is
unlike classical Cosserat dynamics where dissipative effects are usually
ignored. In addition, active forces and torques with no equilibrium
analogue drive the translational and rotational behavior of these
particles. Not only are these forces and torques dependent on both
position and orientation, but also inject power into the surrounding
solvent and thereby may be non-reciprocal and irreversible \citep{finlayson1969}.
By contrast, only potential interactions are admitted in classical
Cosserat dynamics which are both reciprocal and conservative. Finally,
the theoretical description of suspensions are most naturally described
in the particle picture, involving coordinates and orientations, whereas
classical Cosserat dynamics is formulated in the continuum picture,
in terms of displacement and orientation fields. 

In this paper, we study a suspension of active particles conceived
as an active Cosserat crystal, comparing and contrasting their behavior
with that of a Cosserat continuum. We present a theory for the overdamped
mechanics of a suspension of spheroidal active particles, treating
the exchange of momentum between particles and solvent consistently
\citep{singh2018generalized} while also respecting the group-valued
character of the orientational equations of motion. Primacy is given
to the kinematic equations that evolve the positions and orientations
of each particle in the suspension, which may be interpreted as the
action of a group on an $n-$fold product of Euclidean space. The
generators of this action, the velocity and angular velocity, are
obtained by balancing forces and torques in the absence of inertia.
A virtual power principle \citep{finlayson1969,solovev2021lagrangian,germain2020method,germain1973method}
is used to categorize the forces and torques as either conservative,
dissipative or active. We obtain linearized equations of motion for
the kinematic evolution with a Jacobian matrix that is determined
by the derivative of the generators with respect to position and orientation.
Using this general linearized kinematic equation, we study the stability
of an initially crystalline arrangement of active spheroid particles
sedimenting under gravity remote from boundaries \citep{crowley1971viscosity,crowley1976clumping,lahiri1997steadily,PhysRevFluids.4.053102,chajwa2020waves}.
We find that one or more of the linear modes are always unstable which
leads to the melting of the active crystal. This result is elucidated
by identifying a Poisson structure within the odd-dimensional configuration
manifold, consisting of a Hamiltonian in addition to conserved quantities
known as Casimir invariants, with coupled orientation-sedimentation
modes playing the role of conjugate momentum to lattice vibrations
\citep{PhysRevLett.71.3043,PhysRevLett.81.2399,olver2000applications,marsden2013introduction}.
In this way, we explicitly construct Poisson brackets which completely
determine dynamical evolution \citep{goldstein2011classical,arnol2013mathematical}.
While this form of Hamilton's equations is not symmetric under time
reversal, reflecting the presence of irreversible forces and torques,
the non-equilibrium potential implied by the Hamiltonian admits a
conceptual simplification of the stability criteria derived directly
from the linear equations of motion. We conclude with a discussion
on how the long-ranged stability of the active Cosserat crystal differs
from that of the active Cosserat medium and how this approach may
be generally useful in overdamped active particle mechanics \citep{hocking_1964,PhysRevLett.108.218104,zottl2013periodic,lushi2015periodic,stark2016swimming,Shelleyetal,tallapragada2019chaotic,chajwa2019kepler,PhysRevLett.124.088003}. 

\section{Overdamped active mechanics\label{sec:ActiveMech}}

We consider $i=1,\ldots,N$ uniaxial rigid particles of mass $m$
and moment of inertia tensor $\boldsymbol{\mathcal{I}}$ whose centers
of mass are located at $\boldsymbol{R}_{i}$ and whose orientations
are specified by the unit vector $\boldsymbol{p}_{i}$. The kinematic
configuration space is the $N-$fold direct product of three-dimensional
Euclidean space $\mathbb{E}^{3}$ (the space of positions) and the
two-sphere $S^{2}$ (the space of orientations). The velocities $\boldsymbol{V}_{i}$
and angular velocities $\boldsymbol{\Omega}_{i}$ determine the evolution
on this configuration manifold through the kinematic equations,

\begin{align}
\dot{\boldsymbol{R}}_{i} & =\mathbf{V}_{i},\quad\dot{\boldsymbol{p}}_{i}=\mathbf{\boldsymbol{\Omega}}_{i}\times\boldsymbol{p}_{i}.\label{eq:kinematic-first}
\end{align}
The velocities and angular velocities are themselves determined by
Newton's equations of motion for linear and angular momentum, \citep{singh2018generalized,PhysRevLett.124.088003}
\begin{align}
m\dot{\boldsymbol{V}}_{i}=\boldsymbol{F}_{i} & ,\quad\boldsymbol{\mathcal{I}}\cdot\dot{\boldsymbol{\Omega}}_{i}+\boldsymbol{\Omega}_{i}\times\boldsymbol{\mathcal{I}}\cdot\boldsymbol{\Omega}_{i}=\boldsymbol{T}_{i}\label{eq:dynamics-first}
\end{align}
where $\boldsymbol{F}_{i}$ and $\boldsymbol{T}_{i}$ are the total
force and torque acting on the $i$-th particle. It is convenient
to classify the forces and torques by their contribution to the balance
of energy. To this end, we introduce the kinetic energy of the system
through the positive-definite quadratic form 

\begin{equation}
K=\frac{1}{2}m\boldsymbol{V}_{i}\cdot\boldsymbol{V}_{i}+\frac{1}{2}\boldsymbol{\Omega}_{i}\cdot\boldsymbol{\mathcal{I}}\cdot\boldsymbol{\Omega}_{i}
\end{equation}
where repeated particle indices are summed over. Multiplying the equation
for linear momentum by $\boldsymbol{V}_{i}$, the equation of angular
momentum by $\boldsymbol{\Omega}_{i}$, and summing over all particles
we obtain an equation relating the rate of change of kinetic energy
to the power expended by the forces and torques,

\begin{equation}
\frac{dK}{dt}=\boldsymbol{F}_{i}\cdot\boldsymbol{V}_{i}+\boldsymbol{T}_{i}\cdot\boldsymbol{\Omega}_{i}.\label{eq:virtual-power}
\end{equation}
Conservative forces and torques are defined to be even under time
reversal and for which the power expended is the total derivative:

\begin{align}
\boldsymbol{F}_{i}^{\mathcal{C}}\cdot\boldsymbol{V}_{i}+\boldsymbol{T}_{i}^{\mathcal{C}}\cdot\boldsymbol{\Omega}_{i} & =-\frac{dU}{dt}.
\end{align}
This implies the existence of a multibody potential energy function
$U=U(\boldsymbol{R}_{1},\dots\boldsymbol{R}_{N},\boldsymbol{p}_{1},\dots\boldsymbol{p}_{N})$
of the positions and orientations which is even under time reversal.
Consistency with the time derivative of this function and the kinematic
equations then implies that the conservative forces and torques are
related to the potential as 

\begin{equation}
\boldsymbol{F}_{i}^{\mathcal{C}}=-\frac{\partial U}{\partial\boldsymbol{R}_{i}},\quad\boldsymbol{T}_{i}^{\mathcal{C}}=-\boldsymbol{p}_{i}\times\frac{\partial U}{\partial\boldsymbol{p}_{i}}.
\end{equation}
Here we confine our attention to mechanical systems in which the dissipative
forces and torques are linear functions of the velocities and angular
velocities

\begin{equation}
\begin{aligned}\boldsymbol{F}_{i}^{\mathcal{D}} & =-(\boldsymbol{\gamma}_{ij}^{TT}\boldsymbol{V}_{j}+\boldsymbol{\gamma}_{ij}^{TR}\boldsymbol{\Omega}_{j})=\frac{\partial\mathcal{R}}{\partial\boldsymbol{V}_{i}},\\
\boldsymbol{T}_{i}^{\mathcal{D}} & =-(\boldsymbol{\gamma}_{ji}^{RT}\boldsymbol{V}_{j}+\boldsymbol{\gamma}_{ij}^{RR}\boldsymbol{\Omega}_{j})=\frac{\partial\mathcal{R}}{\partial\boldsymbol{\Omega}_{i}}.
\end{aligned}
\end{equation}
Power dissipation is then a positive-definite quadratic form of the
velocities and angular velocities determined by the Rayleigh dissipation
function
\begin{equation}
\boldsymbol{\gamma}_{ij}^{TT}\boldsymbol{V}_{i}\boldsymbol{V}_{j}+2\boldsymbol{\gamma}_{ij}^{TR}\boldsymbol{V}_{i}\boldsymbol{\Omega}_{j}+\boldsymbol{\gamma}_{ij}^{RT}\boldsymbol{\Omega}_{i}\boldsymbol{\Omega}_{j}=2\mathcal{R}>0.
\end{equation}
The matrices $\boldsymbol{\gamma}_{ij}^{TT},\boldsymbol{\gamma}_{ij}^{TR},\boldsymbol{\gamma}_{ij}^{RR}$
are symmetric and positive-definite friction tensors which determine
the dissipative forces and torques. These, in general, are many-body
functions of the positions and orientations of each particle. For
a particle with three planes of symmetry, the resistance tensors obey
the constraint \citep{brenner1963}
\begin{align}
\boldsymbol{\gamma}_{ij}^{TR} & =\boldsymbol{\gamma}_{ij}^{RT},\quad\boldsymbol{\gamma}_{ii}^{TR}=\boldsymbol{\gamma}_{ii}^{RT}=0\label{eq:RTmobilities}
\end{align}
where no sum is implied on the repeated $i$ index. Considering the
remainder of the forces and torques in Eq.(\ref{eq:virtual-power})
to be non-conservative, odd under time reversal and power injecting
leads to the following energy balance equation:

\begin{equation}
\frac{d}{dt}\left(K+U\right)=-2\mathcal{R}+\boldsymbol{F}_{i}^{\mathcal{A}}\cdot\boldsymbol{V}_{i}+\boldsymbol{T}_{i}^{\mathcal{A}}\cdot\boldsymbol{\Omega}_{i}.\label{eq:potential-active-balance}
\end{equation}
This identification of the active forces, $\boldsymbol{F}_{i}^{\mathcal{A}}$,
and active torques, $\boldsymbol{T}_{i}^{\mathcal{A}}$, is similar
to the definition proposed by Finlayson and Scriven, where active
Cauchy stresses in fluid mechanical continua are identified through
a power principle \citep{finlayson1969}. 

When the rate of change of kinetic energy is negligible in the power
balance, the dynamics become overdamped and the rate of change of
potential energy is balanced by dissipation and active power injection.
In this limit, the velocities and angular velocities can be obtained
directly from the simultaneous solution of the momentum and angular
momentum balance equations, 

\begin{equation}
\begin{array}{cc}
\boldsymbol{V}_{i} & =\begin{aligned}\boldsymbol{\mu}_{ij}^{TT}\cdot\left(\boldsymbol{F}_{j}^{\mathcal{C}}+\boldsymbol{F}_{j}^{\mathcal{A}}\right)+\boldsymbol{\mu}_{ij}^{TR}\cdot\left(\boldsymbol{T}_{j}^{\mathcal{C}}+\boldsymbol{T}_{j}^{\mathcal{A}}\right)\end{aligned}
,\\
\boldsymbol{\Omega}_{i} & =\boldsymbol{\mu}_{ij}^{RT}\cdot\left(\boldsymbol{F}_{j}^{\mathcal{C}}+\boldsymbol{F}_{j}^{\mathcal{A}}\right)+\boldsymbol{\mu}_{ij}^{RR}\cdot\left(\boldsymbol{T}_{j}^{\mathcal{C}}+\boldsymbol{T}_{j}^{\mathcal{A}}\right).
\end{array}\label{eq:kinematic}
\end{equation}
The mobility tensors$\boldsymbol{\mu}_{ij}^{TT},\boldsymbol{\mu}_{ij}^{TR},\boldsymbol{\mu}_{ij}^{RR}$
are inverse to the friction tensors introduced above and therefore
inherit identical symmetry properties. The kinematic equations are
closed by the above relations and provide the time-evolution of the
overdamped mechanical system. 

For a passive mechanical system ($\boldsymbol{F}_{i}^{\mathcal{A}}=0,\boldsymbol{T}_{i}^{\mathcal{A}}=0$)
in unbounded potentials, it remains possible that dissipation is balanced
by loss in potential energy resulting in a dynamical steady state.
However, for a potential bounded from below, a dynamical fixed point
is reached with zero velocity and angular velocity, and vanishing
conservative forces and torques. In contrast, for an active mechanical
system ($\boldsymbol{F}_{i}^{\mathcal{A}}\neq0,\boldsymbol{T}_{i}^{\mathcal{A}}\neq0$),
even in the presence of a potential bounded from below, a dynamical
steady state can be reached where the rate of change of potential
energy is balanced by dissipation and the active injection of power.
These steady states could either be fixed points or limit cycles of
the overdamped dynamical system. Much of the surprising aspects of
active mechanical systems can be traced to this property, as will
be apparent in what follows.

\section{Linearized kinematics\label{sec:Linear-stability-of-lattice}}

We now consider active particles with initial configuration $\left(\boldsymbol{R}_{i}^{*},\,\boldsymbol{p}_{i}^{*}\right)$
where $\boldsymbol{R}_{i}^{*}$ is a lattice vector and $\boldsymbol{p}_{i}^{*}$
is a constant vector. Perturbations about this state take the form

\begin{equation}
\boldsymbol{R}_{i}=\boldsymbol{R}_{i}^{\ast}+\boldsymbol{u}_{i},\quad\boldsymbol{p}_{i}=\boldsymbol{p}_{i}^{\ast}+\boldsymbol{q}_{i}.\label{eq:steady state}
\end{equation}
Expanding the velocity and angular velocity to first order in these
perturbations yields

\begin{equation}
\begin{array}{cc}
\boldsymbol{V}_{i}= & \boldsymbol{V}_{i}^{\ast}+\left[\frac{\partial\boldsymbol{V}_{i}}{\partial\boldsymbol{R}_{j}}\right]_{\ast}\cdot\boldsymbol{u}_{j}+\left[\frac{\partial\boldsymbol{V}_{i}}{\partial\boldsymbol{p}_{j}}\right]_{\ast}\cdot\boldsymbol{q}_{j},\\
\boldsymbol{\Omega}_{i}= & \boldsymbol{\Omega}_{i}^{\ast}+\left[\frac{\partial\boldsymbol{\Omega}_{i}}{\partial\boldsymbol{R}_{j}}\right]_{\ast}\cdot\boldsymbol{u}_{j}+\left[\frac{\partial\boldsymbol{\Omega}_{i}}{\partial\boldsymbol{p}_{j}}\right]_{\ast}\cdot\boldsymbol{q}_{j}.
\end{array}
\end{equation}
Inserting this into Eq.(\ref{eq:kinematic-first}) yields
\begin{equation}
\begin{aligned}\dot{\boldsymbol{u}}_{i}= & \boldsymbol{V}_{i}^{\ast}+\left[\frac{\partial\boldsymbol{V}_{i}}{\partial\boldsymbol{R}_{j}}\right]_{\ast}\cdot\boldsymbol{u}_{j}+\left[\frac{\partial\boldsymbol{V}_{i}}{\partial\boldsymbol{p}_{j}}\right]_{\ast}\cdot\boldsymbol{q}_{j},\\
\dot{\boldsymbol{q}}_{i}= & \boldsymbol{\Omega}_{i}^{\ast}\times\boldsymbol{p}_{i}^{*}+\left[\boldsymbol{p}_{i}\times\frac{\partial\boldsymbol{\Omega}_{i}}{\partial\boldsymbol{R}_{j}}\right]_{*}\cdot\boldsymbol{u}_{j}\\
+ & \boldsymbol{\Omega}_{i}^{\ast}\times\boldsymbol{q}_{i}+\,\left[\boldsymbol{p}_{i}\times\frac{\partial\boldsymbol{\Omega}_{i}}{\partial\boldsymbol{p}_{j}}\right]_{\ast}\cdot\boldsymbol{q}_{j},
\end{aligned}
\label{eq:linear-1st-order}
\end{equation}
where the zeroth order terms $\boldsymbol{V}_{i}^{\ast},\,\boldsymbol{\Omega}_{i}^{\ast}\times\boldsymbol{p}_{i}^{*}$
represent rigid body motions of the initial state. When $\boldsymbol{\Omega}_{i}^{\ast}\times\boldsymbol{p}_{i}^{*}=0$
and the $\boldsymbol{V}_{i}^{*}$ are the same for every particle,
we may transform to a co-moving frame where the dynamics are defined
by a $6N\times6N$ linear system of $3N$ translational and $3N$
rotational degrees of freedom
\begin{equation}
\frac{d}{dt}\begin{pmatrix}\boldsymbol{u}_{i}\\
\boldsymbol{q}_{i}
\end{pmatrix}=\begin{pmatrix}\boldsymbol{J}^{uu} & \boldsymbol{J}^{uq}\\
\boldsymbol{J}^{qu} & \boldsymbol{J}^{qq}
\end{pmatrix}_{ij}\begin{pmatrix}\boldsymbol{u}_{j}\\
\boldsymbol{q}_{j}
\end{pmatrix}\label{eq:linear-evolution}
\end{equation}
which describes the linear evolution of perturbations with $3N\times3N$
block Jacobian matrices taking the form

\begin{equation}
\begin{aligned}\boldsymbol{J}_{ij}^{uu}= & \frac{\partial\boldsymbol{V}_{i}}{\partial\boldsymbol{R}_{j}}, & \boldsymbol{J}_{ij}^{uq}= & \frac{\partial\boldsymbol{V}_{i}}{\partial\boldsymbol{p}_{j}},\\
\boldsymbol{J}_{ij}^{qu}= & \boldsymbol{p}_{i}^{*}\times\frac{\partial\boldsymbol{\Omega}_{i}}{\partial\boldsymbol{R}_{j}}, & \boldsymbol{J}_{ij}^{qq}= & \boldsymbol{p}_{i}^{*}\times\frac{\partial\boldsymbol{\Omega}_{i}}{\partial\boldsymbol{p}_{j}}.
\end{aligned}
\end{equation}
For the overdamped dynamics, these Jacobian matrices can be obtained
from the differentials of the velocities and angular velocities

\begin{equation}
\begin{aligned}d\boldsymbol{V}_{i} & =\begin{aligned}d\begin{aligned}\boldsymbol{\mu}_{ij}^{TT}\cdot\left(\boldsymbol{F}_{j}^{\mathcal{C}}+\boldsymbol{F}_{j}^{\mathcal{A}}\right)+d\boldsymbol{\mu}_{ij}^{TR}\cdot\left(\boldsymbol{T}_{j}^{\mathcal{C}}+\boldsymbol{T}_{j}^{\mathcal{A}}\right)\end{aligned}
\end{aligned}
\\
 & +\begin{aligned}\begin{aligned}\boldsymbol{\mu}_{ij}^{TT}\cdot d\left(\boldsymbol{F}_{j}^{\mathcal{C}}+\boldsymbol{F}_{j}^{\mathcal{A}}\right)+\boldsymbol{\mu}_{ij}^{TR}\cdot d\left(\boldsymbol{T}_{j}^{\mathcal{C}}+\boldsymbol{T}_{j}^{\mathcal{A}}\right),\end{aligned}
\end{aligned}
\\
d\boldsymbol{\Omega}_{i} & =\begin{aligned}d\begin{aligned}\boldsymbol{\mu}_{ij}^{RT}\cdot\left(\boldsymbol{F}_{j}^{\mathcal{C}}+\boldsymbol{F}_{j}^{\mathcal{A}}\right)+d\boldsymbol{\mu}_{ij}^{RR}\cdot\left(\boldsymbol{T}_{j}^{\mathcal{C}}+\boldsymbol{T}_{j}^{\mathcal{A}}\right)\end{aligned}
\end{aligned}
\\
 & +\begin{aligned}\begin{aligned}\boldsymbol{\mu}_{ij}^{RT}\cdot d\left(\boldsymbol{F}_{j}^{\mathcal{C}}+\boldsymbol{F}_{j}^{\mathcal{A}}\right)+\boldsymbol{\mu}_{ij}^{RR}\cdot d\left(\boldsymbol{T}_{j}^{\mathcal{C}}+\boldsymbol{T}_{j}^{\mathcal{A}}\right).\end{aligned}
\end{aligned}
\end{aligned}
\label{eq:linear-J-1}
\end{equation}
These differentials receive contributions from the infinitesimal changes
in the mobilities and the infinitesimal changes in the forces and
torques as the configurations $\boldsymbol{R}_{i},\boldsymbol{p}_{i}$
are varied. It is important to note that the differentials of the
configurations must obey the $N$ kinematic constraints $\boldsymbol{p}_{i}\cdot\dot{\boldsymbol{p}}_{i}=\boldsymbol{p}_{i}\cdot\left(\boldsymbol{\Omega}_{i}\times\boldsymbol{p}_{i}\right)=0$.
This reflects the fact that $\boldsymbol{p}_{i}$ is a coordinate
on $S^{2}$ requiring $\boldsymbol{p}_{i}^{*}\cdot\boldsymbol{q}_{i}=0$.
Orientational fluctuations are therefore described by the remaining
two degrees of freedom in $\boldsymbol{q}_{i}$. In particular, the
differential $d\boldsymbol{p}$ must be perpendicular to $\boldsymbol{p}_{i}^{\ast}$
and thus for a parametrization $\boldsymbol{p}_{i}^{\ast}=(0,0,1)$
this implies $d\boldsymbol{p}=(dp_{x},dp_{y},0)$. In the following
section we specify the forms of the mobilities and the forces and
torques for a suspension of hydrodynamically interacting active particles. 

\section{Active forces and torques\label{sec:Hydrodynamic-interactions}}

The most common experimental realization of an overdamped active mechanical
system, in the sense of Sec. (\ref{sec:ActiveMech}), is a suspension
of active colloids \citep{mognetti2013livingClusters,herminghaus2014interfacial,maass2016swimming,Kruger2016,Seemann2016,caciagli2020controlled}.
Active colloids are endowed with microscopic mechanisms that generate
slip velocities on the particle-fluid boundaries. An expansion of
the slip that yields self-propulsion and self-rotation takes the form
\citep{singh2018generalized}
\begin{equation}
\boldsymbol{v}_{i}^{A}=\boldsymbol{V}_{i}^{\mathcal{A}}+\boldsymbol{\Omega}_{i}^{\mathcal{A}}\times\boldsymbol{\rho}_{i}
\end{equation}
where $\boldsymbol{\rho}_{i}$ is a vector parameterizing the surface
of the $i$-th particle. The coefficients $\boldsymbol{V}_{i}^{\mathcal{A}}$
and $\boldsymbol{\Omega}_{i}^{\mathcal{A}}$ are here taken to be
functions of the orientation of the $i$-th particle and will be specified
below. The slips produces stresses in the fluid that act back on the
particles as tractions on the particle-fluid boundaries. The active
forces and torques are obtained by integrating 
\begin{equation}
\boldsymbol{F}_{i}^{\mathcal{A}}=\int\boldsymbol{t}_{i}^{\mathcal{A}}dS_{i},\quad\boldsymbol{T}_{i}^{\mathcal{A}}=\int\boldsymbol{\rho}_{i}\times\boldsymbol{t}_{i}^{\mathcal{A}}dS_{i}\label{eq:boundaryInts}
\end{equation}
where $\boldsymbol{t}_{i}^{\mathcal{A}}$ is the slip-dependent traction
on the $i$-th particle due to its own activity and that of all the
remaining particles. The linearity of the Stokes equations and of
the boundary conditions implies that the active forces and torques
are of the form \citep{singh2018generalized,singh2018microhydrodynamics}
\begin{equation}
\begin{aligned}\boldsymbol{F}_{i}^{\mathcal{A}} & =\boldsymbol{\gamma}_{ij}^{TT}\cdot\boldsymbol{V}_{j}^{\mathcal{A}}+\boldsymbol{\gamma}_{ij}^{TR}\cdot\boldsymbol{\Omega}_{j}^{\mathcal{A}},\\
\boldsymbol{T}_{i}^{\mathcal{A}} & =\boldsymbol{\gamma}_{ij}^{RT}\cdot\boldsymbol{V}_{j}^{\mathcal{A}}+\boldsymbol{\gamma}_{ij}^{RR}\cdot\boldsymbol{\Omega}_{j}^{\mathcal{A}}
\end{aligned}
\label{eq:active-forces}
\end{equation}
which represent a generalization of Stokes laws of friction for active
colloids. This specifies the velocities and angular velocities of
the particles in terms of their configuration in a manner that is
consistent with the conservation of momentum and angular momentum
in the suspension.

The hydrodynamic self-mobilities for a body with uniaxial spheroidal
symmetry must be of the form \citep{brenner1963,kim2005} 
\begin{align}
\boldsymbol{\mu}_{ii}^{TT} & =\mu_{1}^{T}\left(\boldsymbol{I}-\boldsymbol{p}_{i}\boldsymbol{p}_{i}\right)+\mu_{2}^{T}\boldsymbol{p}_{i}\boldsymbol{p}_{i},\nonumber \\
\boldsymbol{\mu}_{ii}^{RR} & =\mu_{1}^{R}\left(\boldsymbol{I}-\boldsymbol{p}_{i}\boldsymbol{p}_{i}\right)+\mu_{2}^{R}\boldsymbol{p}_{i}\boldsymbol{p}_{i},\label{eq:uniaxial-one-body-mobility-geom}\\
\boldsymbol{\mu}_{ii}^{TR} & =\boldsymbol{\mu}_{ii}^{RT}=0\nonumber 
\end{align}
where no summation is implied on the $i$ index. The scalar coefficients
$\mu_{1}^{T}\neq\mu_{2}^{T}$ and $\mu_{1}^{R}\neq\mu_{2}^{R}$ reflect
the anisotropy of the translational and rotational responses along
and perpendicular to the axis of uniaxial symmetry. There is no hydrodynamic
coupling between translation and rotation in the absence of chirality,
as we have noted before in Eq.(\ref{eq:RTmobilities}). The mutual
mobilities in the pair approximation are given by 
\begin{equation}
\begin{aligned}\boldsymbol{\mu}_{ij}^{TT}= & \mathcal{F}_{i}^{0}\left(\boldsymbol{p}_{i}\right)\mathcal{F}_{j}^{0}\left(\boldsymbol{p}_{j}\right)\boldsymbol{G}_{ij}\left(\boldsymbol{R}_{i},\boldsymbol{R}_{j}\right),\\
\boldsymbol{\mu}_{ij}^{RT}= & \frac{1}{2}\mathcal{F}_{i}^{1}\left(\boldsymbol{p}_{i}\right)\mathcal{F}_{j}^{0}\left(\boldsymbol{p}_{j}\right)\boldsymbol{\nabla}_{\boldsymbol{R}_{i}}\times\boldsymbol{G}_{ij}\left(\boldsymbol{R}_{i},\boldsymbol{R}_{j}\right),
\end{aligned}
\end{equation}
where $\boldsymbol{G}_{ij}$ is a Green's function of the Stokes equation
that determines the bulk fluid flow resulting from a point force,
while $\mathcal{F}_{i}^{0}$ and $\mathcal{F}_{i}^{1}$ are Faxén
operators that correct for the finite size of the particles \citep{faxen1922widerstand,kim2005}.
In the dilute limit, the leading contributions to the mutual mobilities
do not depend on the finiteness of the particles and the Faxén operators
may be set to the identity. The Green's function for a linear medium
of infinite extent is given by
\begin{equation}
\boldsymbol{G}\left(\boldsymbol{R}_{i},\boldsymbol{R}_{j}\right)=A_{1}\boldsymbol{I}+A_{2}\hat{\boldsymbol{r}}_{ij}\hat{\boldsymbol{r}}_{ij},\label{eq:greens-function}
\end{equation}
where $\boldsymbol{r}_{ij}=\boldsymbol{R}_{i}-\boldsymbol{R}_{j}$
and $\hat{\boldsymbol{r}}_{ij}=\boldsymbol{r}_{ij}/r_{ij}$ is the
normalized separation between particles $i$ and $j$. For a viscous
overdamped Stokes medium, the Oseen tensor is defined with the coefficients
\begin{equation}
A_{1}=A_{2}=\frac{1}{8\pi\eta r_{ij}},
\end{equation}
where $\eta$ is the viscosity of the external fluid. Specific choices
of the active velocities and angular velocities $\boldsymbol{V}_{A},\boldsymbol{\Omega}_{A}$
along with conservative forces and torques $\boldsymbol{F}^{\mathcal{C}},\boldsymbol{T}^{\mathcal{C}}$
can now be made to study particular overdamped active systems. 

\section{sedimenting active crystals\label{sec:sedimenting-active-crystals}}

We now investigate the linear stability of a lattice of identical
active particles sedimenting under gravity in a Stokesian fluid of
infinite extent. This idealizes relevant experiment conditions \citep{crowley1971viscosity,crowley1976clumping,chajwa2020waves}
where active particles remain remote from the boundaries of the container.
The conservative and active forces are determined by the expressions

\begin{equation}
\begin{aligned}\boldsymbol{F}_{i}^{\mathcal{C}}= & m\boldsymbol{g},\quad\boldsymbol{T}_{i}^{\mathcal{C}}=0,\\
\boldsymbol{V}_{i}^{\mathcal{A}}= & v_{A}\boldsymbol{p}_{i},\quad\boldsymbol{\Omega}_{i}^{\mathcal{A}}=\omega_{A}\boldsymbol{p}_{i},
\end{aligned}
\end{equation}
 where $m$ is the buoyant mass of the particle, $v_{A}$ is the active
speed and $\omega_{A}$ is the active angular speed. The rigid body
motion of the particles is then determined by the pair of equations
\begin{equation}
\begin{aligned}\boldsymbol{V}_{i} & =\begin{aligned}\sum_{j}\boldsymbol{\mu}_{ij}^{TT}\cdot m\boldsymbol{g}+v_{A}\boldsymbol{p}_{i}\end{aligned}
,\\
\boldsymbol{\Omega}_{i} & =\sum_{j\neq i}\boldsymbol{\mu}_{ij}^{RT}\cdot m\boldsymbol{g}+\omega_{A}\boldsymbol{p}_{i},
\end{aligned}
\label{eq:kinematic-sed-sub}
\end{equation}
representing the one-body translational motion of the $i$-th particle
under gravitational force and active motion, and hydrodynamic two-body
interactions due to entrainment in the flow field produced by the
other sedimenting particles. The rotational motion of the $i$-th
particle has no one-body hydrodynamic contribution and reorientation
results entirely from the vorticity produced by the sedimentation
flow. These equations generalize the dynamics of active particle pairs
presented in \citep{PhysRevLett.124.088003} to a many-body system.
The elements of the Jacobian matrix that follow from the above are
\begin{equation}
\begin{aligned}\boldsymbol{J}_{ij}^{uu} & =\sum_{l}\left[\frac{\partial\boldsymbol{\mu}_{il}^{TT}}{\partial\boldsymbol{R}_{j}}\cdot m\boldsymbol{g}\right]_{*},\\
\boldsymbol{J}_{ij}^{uq} & =\sum_{l}\left[\frac{\partial\boldsymbol{\mu}_{il}^{TT}}{\partial\boldsymbol{p}_{j}}\cdot m\boldsymbol{g}\right]_{*}+v_{A}\delta_{ij}\text{diag}\left(1,1,0\right),\\
\boldsymbol{J}_{ij}^{qu} & =\sum_{l}\left[\boldsymbol{p}_{i}\times\frac{\partial\boldsymbol{\mu}_{il}^{RT}}{\partial\boldsymbol{R}_{j}}\cdot m\boldsymbol{g}\right]_{*},\\
\boldsymbol{J}_{ij}^{qq} & =0.
\end{aligned}
\label{eq:1dchain-sum}
\end{equation}
In the following sections we shall use these to analyze the stability
of one- and two-dimensional lattice. 

We conclude this section with an observation about the contributions
from geometric anisotropy and activity to translational motion. Combining
Eqs.(\ref{eq:uniaxial-one-body-mobility-geom},\ref{eq:kinematic-sed-sub}),
the velocity of the $i-$th particle may be expressed as 
\begin{equation}
\boldsymbol{V}_{i}=\mu_{1}^{T}m\boldsymbol{g}-\left[\left(\mu_{1}^{T}-\mu_{2}^{T}\right)\left(\boldsymbol{p}_{i}\cdot m\boldsymbol{g}\right)-v_{A}\right]\boldsymbol{p}_{i}+\text{HI}
\end{equation}
where the contributions to the one-body mobility from geometric anisotropy
and activity have been made clear and where HI includes mutual hydrodynamic
contributions. Here, geometric anisotropy adds an apolar term to the
velocity while activity adds a polar term. For small perturbation,
however, the contribution takes the form 
\begin{equation}
\boldsymbol{J}_{ii}^{uq}\cdot\boldsymbol{q}_{i}=\left[\left(\mu_{1}^{T}-\mu_{2}^{T}\right)mg+v_{A}\right]\boldsymbol{q}_{i}
\end{equation}
showing that it is not possible to distinguish geometric anisotropy
from activity at the level of the linearized equations of motion.
We may therefore define an effective active speed
\begin{equation}
\tilde{v}=\left(\mu_{1}^{T}-\mu_{2}^{T}\right)mg+v_{A}
\end{equation}
encapsulating both the passive effect of geometric anisotropy and
the active effect of self-propulsion as the single relevant parameter
controlling the dynamics. In what follows, we shall assume spherical
particles of radius $b$ with $\mu_{1}^{T}=\mu_{2}^{T}=\mu^{T}=6\pi\eta b$
with a self-propulsion speed $v_{A}$, with the understanding that
this causes no loss of generality.

\section{One-dimensional lattice\label{sec:Sedimenting-active-1D}}

\begin{figure}
\includegraphics[width=1\columnwidth]{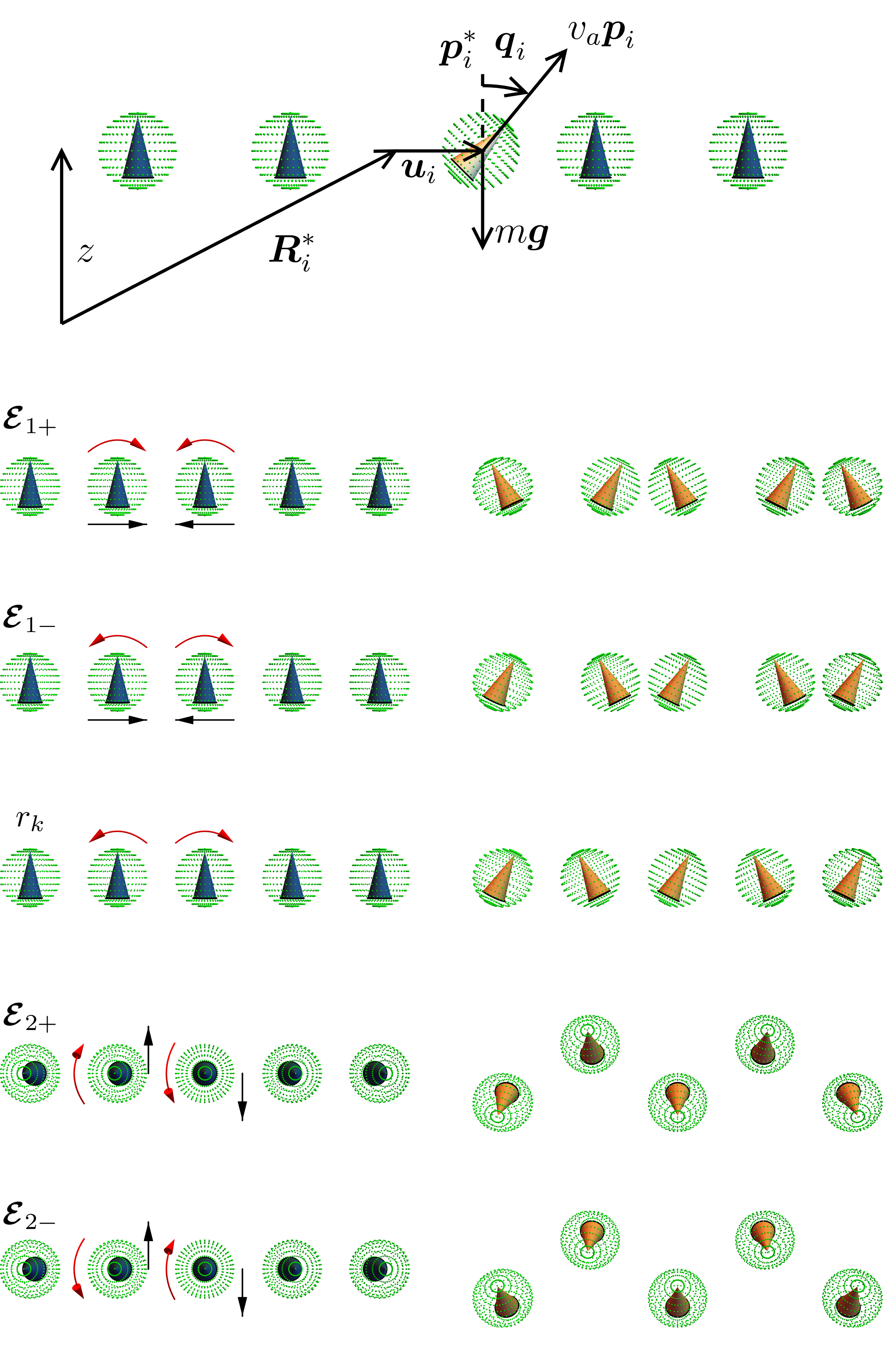}

\caption{\label{fig:Schematic-of1dchain}The first panel shows a schematic
of the sedimenting one-dimensional crystal as specified by Eqs.(\ref{eq:linear-1st-order},\ref{eq:initial-config-1d}).
The remaining panels present a graphical depiction of the collective
modes presented in Tab.(\ref{tab:eigensystem-1d}) evaluated at wavenumber
$k=\pi/2$. The first three modes lie in the space co-ordinatized
by the triplet $\text{\ensuremath{\left(u_{1k},q_{1k},u_{3k}\right)}}$
and are viewed perpendicular to the direction of gravity while the
remaining two modes lie in the space co-ordinatized by the pair $\left(u_{2k},q_{2k}\right)$
and are viewed along the direction of gravity.}
\end{figure}

We now consider a one-dimensional lattice of particles with initial
positions and orientations 

\begin{equation}
\boldsymbol{R}_{i}^{*}=\left(x_{i}^{*},0,0\right),\quad\boldsymbol{p}_{i}^{*}=\left(0,0,1\right)\label{eq:initial-config-1d}
\end{equation}
where $x_{i}^{*}=ia$ and $a$ is the lattice spacing. The discrete
translational invariance of the system can be exploited to diagonalize
the linearized dynamics in terms of plane wave collective modes

\[
\boldsymbol{u}_{i}\left(t\right)=b\int\frac{dk}{2\pi}e^{ikx_{i}^{*}}\boldsymbol{u}_{k}\left(t\right),\quad\boldsymbol{q}_{i}\left(t\right)=\int\frac{dk}{2\pi}e^{ikx_{i}^{*}}\boldsymbol{q}_{k}\left(t\right),
\]
where the wave vector $\boldsymbol{k}=\left(k,0,0\right)$ is directed
along the chain. Inserting these plane wave solutions into Eq.(\ref{eq:linear-evolution})
and summing over all lattice points yields the Fourier-transformed
linear equations of motion

\begin{equation}
\frac{d}{dt}\begin{pmatrix}\boldsymbol{u}_{k}\\
\boldsymbol{q}_{k}
\end{pmatrix}=\begin{pmatrix}\boldsymbol{J}^{uu} & \boldsymbol{J}^{uq}\\
\boldsymbol{J}^{qu} & \boldsymbol{J}^{qq}
\end{pmatrix}_{k}\begin{pmatrix}\boldsymbol{u}_{k}\\
\boldsymbol{q}_{k}
\end{pmatrix},\label{eq:linear-evolution-kspace}
\end{equation}
where the block Jacobian elements are given by

\[
\boldsymbol{J}_{k}^{\alpha\beta}=\sum_{\text{lattice}}\boldsymbol{J}_{ij}^{\alpha\beta}e^{-ik\left(x_{i}^{*}-x_{j}^{*}\right)}.
\]
with $\alpha,\beta\in\left\{ u,q\right\} $. A schematic of the initial
condition and kinematic variables is presented in Fig.(\ref{fig:Schematic-of1dchain}).
Introducing the Stokes velocity $v_{0}$, the Stokes time $\tau$
and the dimensionless active speed $v$ through the relations 
\[
v_{0}=\mu^{T}mg,\quad\tau=b/v_{0},\quad v=\tilde{v}/v_{0}
\]
and evaluating the required matrix elements (detailed in Appendix A),
the Jacobian matrix in the plane-wave basis is 

\begin{equation}
\boldsymbol{J}_{k}=\left(\begin{array}{ccc|ccc}
0 & 0 & -s\left(k\right) & v & 0 & 0\\
0 & 0 & 0 & 0 & v & 0\\
s\left(k\right) & 0 & 0 & 0 & 0 & 0\\
\hline -2c\left(k\right) & 0 & 0 & 0 & 0 & 0\\
0 & c\left(k\right) & 0 & 0 & 0 & 0\\
0 & 0 & 0 & 0 & 0 & 0
\end{array}\right),\label{eq:1dJk}
\end{equation}
where
\begin{equation}
\begin{alignedat}{1}\lambda & =\frac{3b^{2}}{2a},\\
s\left(k\right) & =\frac{i\lambda}{a}\sum_{n=1}^{\infty}\frac{\sin\left(nka\right)}{n^{2}}\equiv s,\\
c\left(k\right) & =\frac{\lambda b}{a^{2}}\sum_{n=1}^{\infty}\frac{\cos\left(nka\right)-1}{n^{3}}\equiv c.
\end{alignedat}
\label{eq:Lfuncs}
\end{equation}
We note that $s\left(k\right)$ is purely imaginary while $c\left(k\right)$
is real and negative. The collective modes consist of a longitudinal
positional polarization $u_{1k}$, a pair of transverse positional
polarizations $u_{2k},u_{3k}$, a longitudinal orientational polarization
$q_{1k}$ and a transverse orientational polarization $q_{2k}$. The
constraint $\boldsymbol{p}^{*}\cdot\boldsymbol{q}_{k}$ eliminates
the third degree of freedom in the orientational collective mode $\boldsymbol{q}_{k}$
so that only five degrees of freedom remain. This is represented by
the vanishing of the final row and column in $\boldsymbol{J}_{k}$
leading to a trivial null eigenvalue. The remaining dynamical degrees
of freedom yield five non-trivial linear equations of motion 
\begin{equation}
\begin{aligned}\dot{u}_{1k} & =-su_{3k}+vq_{1k},\quad\dot{u}_{2k}=vq_{2k},\quad\dot{u}_{3k}=su_{1k},\\
\dot{q}_{1k} & =-2cu_{1k},\quad\dot{q}_{2k}=cu_{2k}.
\end{aligned}
\label{eq:explicit-1d-eqns}
\end{equation}
These equations show that the couplings between the positional and
orientational collective modes are not symmetric, a property that
can be traced to the non-potential character of the active forces
and torques. The linear system decouples into two subspaces. The first
subspace consists of the triplet $u_{1k},q_{1k},u_{3k}$ which are
a pair of longitudinal modes and the transverse mode parallel to gravity.
These obey the linear equations
\begin{equation}
\frac{d}{dt}\begin{pmatrix}u_{1k}\\
q_{1k}\\
u_{3k}
\end{pmatrix}=\begin{pmatrix}0 & v & -s\\
-2c & 0 & 0\\
s & 0 & 0
\end{pmatrix}\begin{pmatrix}u_{1k}\\
q_{1k}\\
u_{3k}
\end{pmatrix}.\label{eq:3dlindisp}
\end{equation}
One may construct a conserved quantity orthogonal to the dynamical
flow given by
\[
\mathcal{E}_{0}=2cu_{3k}+sq_{1k}\equiv r_{k},\quad\dot{r}_{k}=0.
\]
This non-trivial conserved quantity reflects the fact that local activity
may work to oppose passive clumping of the lattice. The dimensionality
of the first subspace is therefore reduced by one, leading to the
appearance of two linearly independent position-orientation-sedimentation
coupled eigenmodes given by 
\begin{equation}
\mathcal{E}_{1\pm}=\pm\sqrt{-s^{2}-2cv}u_{1k}+vq_{1k}-su_{3k}.
\end{equation}
The positive sign denotes an in-phase longitudinal position and orientation
wave while the negative sign denotes its anti-phase counterpart. These
eigenmodes obey the harmonic equation
\begin{equation}
\ddot{\mathcal{E}}_{1\pm}=-\omega_{1}^{2}\mathcal{E}_{1\pm},\,\omega_{1}^{2}=s^{2}+2cv.
\end{equation}
This immediately implies the existence of stable harmonic waves of
frequency $\omega_{1}$ when $v<-s^{2}/2c$, which is negative, such
that $\omega_{1}^{2}>0$ \citep{chajwa2020waves}. Conversely, when
$v>-s^{2}/2c$ the square of the frequency $\omega_{1}^{2}<0$ leads
to the presence of an exponentially growing mode $\mathcal{E}_{1+}$
while the conjugate mode $\mathcal{E}_{1-}$ is exponentially decaying.
The solution for passive spheres is obtained by taking $v=0$ which
immediately implies that passive sedimenting one-dimensional lattices
are unstable \citep{crowley1971viscosity}. The second subspace consists
of the pair of transverse modes $u_{2k},q_{2k}$ which obey the linear
equations
\begin{equation}
\frac{d}{dt}\begin{pmatrix}u_{2k}\\
q_{2k}
\end{pmatrix}=\begin{pmatrix}0 & v\\
c & 0
\end{pmatrix}\begin{pmatrix}u_{2k}\\
q_{2k}
\end{pmatrix}.
\end{equation}
This leads to the appearance of two linearly independent transverse
position-orientation coupled eigenmodes given by
\begin{equation}
\mathcal{E}_{2\pm}=\pm\sqrt{cv}u_{2k}+vq_{2k}.
\end{equation}
Again, the positive sign denotes an in-phase transverse position and
orientation wave while the negative sign denotes its anti-phase counterpart.
These eigenmodes obey a different harmonic equation given by
\begin{equation}
\ddot{\mathcal{E}}_{2\pm}=-\omega_{2}^{2}\mathcal{E}_{2\pm},\,\omega_{2}^{2}=-cv.
\end{equation}
In this case, for $v>0$ we obtain wavelike solutions of frequency
$\omega_{2}$. However, for $v<0$ we see that the square of the frequency
$\omega_{2}^{2}<0$ leads to the presence of an exponentially growing
mode $\mathcal{E}_{2+}$ while the conjugate mode $\mathcal{E}_{2-}$
is exponentially decaying. For passive spheres, $v=\omega_{2}=0$
such that transverse perturbations are neither stable nor unstable.
Together, we see that there is no value of $v$ for which both longitudinal
and transverse modes are stable, and thus an exponentially growing
solution will always be present. Furthermore, for the range of parameter
values $0>v>-s^{2}/2c$ both transverse and longitudinal modes $\mathcal{E}_{1+},\mathcal{E}_{2+}$
are unstable. 
 
\begin{table*}
\begin{tabular}{|c|c|c|}
\hline 
Eigenvalue $\Lambda_{\alpha}$ & Eigenvector $\boldsymbol{\mathcal{E}}_{\alpha}$ & Interpretation\tabularnewline
\hline 
\hline 
$\Lambda_{0}=0$ & $r_{k}=2cu_{3k}+sq_{1k}$ & one-body orientation coupled to sedimentation velocity\tabularnewline
\hline 
$\Lambda_{1\pm}=\pm\omega_{1}$ & $\mathcal{E}_{1\pm}=\pm\sqrt{-s^{2}-2cv}u_{1k}+vq_{1k}-su_{3k}$ & coupled in-/anti-phase longitudinal position-orientation mode\tabularnewline
\hline 
$\Lambda_{2\pm}=\pm\omega_{2}$ & $\mathcal{E}_{2\pm}=\pm\sqrt{cv}u_{2k}+vq_{2k}$ & coupled in-/anti-phase transverse position-orientation mode\tabularnewline
\hline 
\end{tabular}

\caption{\label{tab:eigensystem-1d}Tabulation of the eigensystem of Eq.(\ref{eq:1dJk})
where $\omega_{1}^{2}=s^{2}+2cv,\quad\omega_{2}^{2}=-cv$.}
\end{table*}

In the following section we show that these equations of motion admit
symplectic structure. This allows us to reformulate the stability
criteria through the construction of a scalar Hamiltonian function,
thus elucidating the stability mechanism of the crystal. 

\subsection{Poisson structure}

\begin{figure}
\includegraphics[width=1\columnwidth]{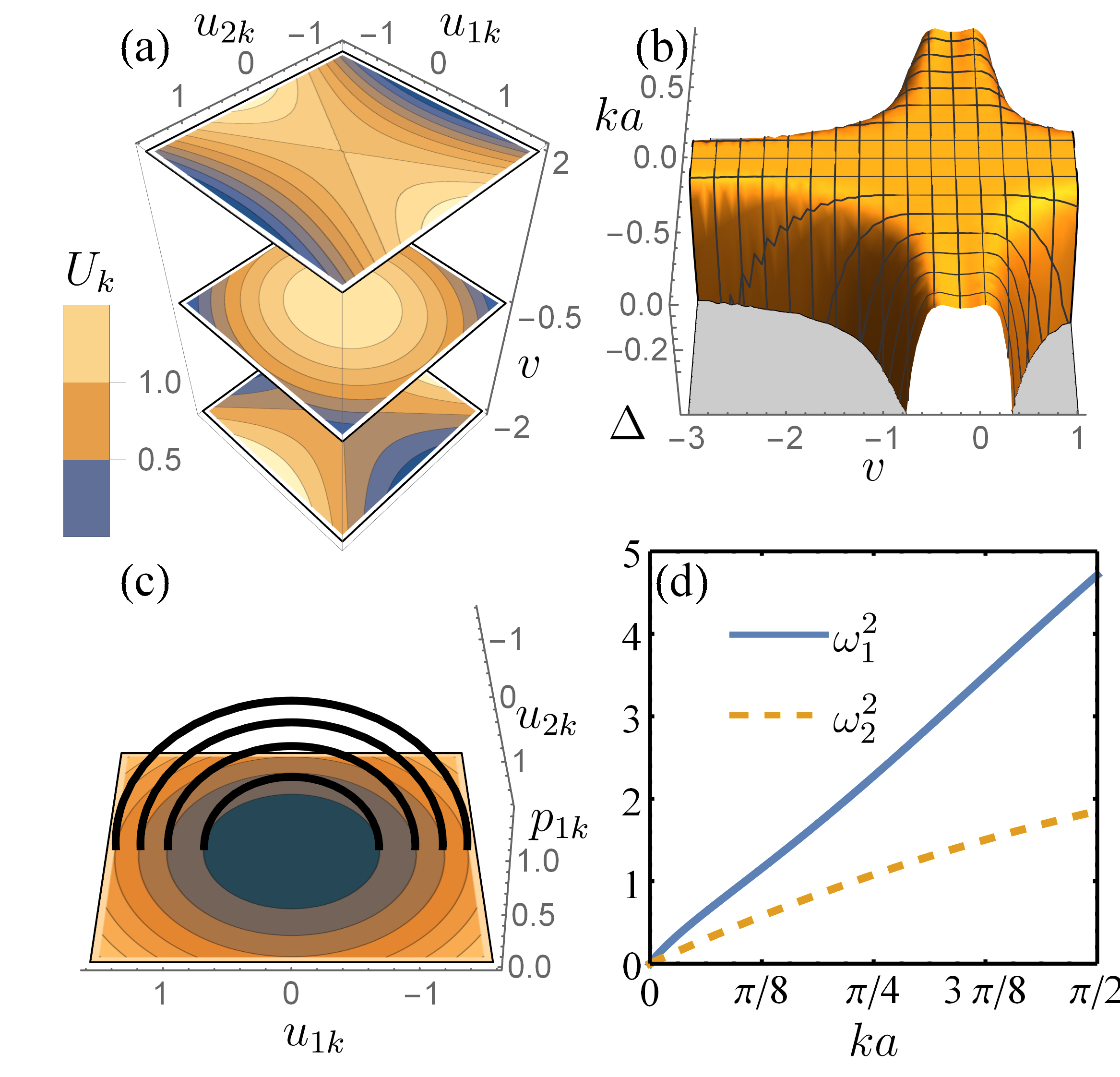}

\caption{\label{fig:Hkpic}(a) Contour plots of the potential $U_{k}$ for
$v=-2,-0.5,2$. For $v=2$ the $u_{1k}$ direction is stable while
the $u_{2k}$ direction is unstable. For $v=-0.5$ both $u_{1k}$
and $u_{2k}$ directions are unstable. For $v=-2$ the $u_{1k}$ direction
is unstable while the $u_{2k}$ direction is stable. (b) Plot of
the Hessian determinant $\Delta=\omega_{1}^{2}\omega_{2}^{2}$ of
$U_{k}$ against $v,ka$. Negative values of $\Delta$ indicate the
presence of a saddle point with a single unstable mode, while positive values of $\Delta$
indicate a parabolic potential with two unstable modes since $\Delta>0$
when $0>v>-s^{2}/2c$. (c) Contour plot of Eq.(\ref{eq:potential})
for anisotropic activity $v_{1}=-1.5,v_{2}=1$ with a superimposed
Poincaré recurrence plot at $u_{2k}=0$. The appearance of closed
one-dimensional contours indicates stable quasiperiodic orbital behavior.
(d) Plot of $\omega_{1}^{2},\omega_{2}^{2}$ against $ka$ for the
same anisotropic activity. These frequencies are positive for all
values of $ka$ leading to a stable positive definite Hamiltonian
with massless dispersion.}
\end{figure}

We start by making the co-ordinate transformations
\begin{align}
p_{1k}=-su_{3k}+vq_{1k},\quad p_{2k}=vq_{2k}
\end{align}
such that the linear evolution can be cast in the form
\begin{align}
\frac{d}{dt}\begin{pmatrix}u_{1k}\\
u_{2k}\\
p_{1k}\\
p_{2k}\\
r_{k}
\end{pmatrix} & =\left(\begin{array}{cccc|c}
0 & 0 & 1 & 0 & 0\\
0 & 0 & 0 & 1 & 0\\
-1 & 0 & 0 & 0 & 0\\
0 & -1 & 0 & 0 & 0\\
\hline 0 & 0 & 0 & 0 & 0
\end{array}\right)\boldsymbol{\nabla}H_{k},\label{eq:poisson-structure}
\end{align}
\[
\boldsymbol{\nabla}=\begin{pmatrix}\partial/\partial u_{1k}, & \partial/\partial u_{2k}, & \partial/\partial p_{1k}, & \partial/\partial p_{2k}, & \partial/\partial r_{k}\end{pmatrix},
\]
with Hamiltonian
\begin{equation}
\begin{array}{cc}
H_{k} & =\frac{1}{2}\left(p_{1k}^{2}+p_{2k}^{2}\right)+\frac{1}{2}\left(\omega_{1}^{2}u_{1k}^{2}+\omega_{2}^{2}u_{2k}^{2}\right),\end{array}\label{eq:hamiltonian-k}
\end{equation}
where $p_{1k},\,p_{2k}$ are momentum-like variables conjugate to longitudinal
and transverse displacements respectively. $p_{1k}$ consists of a
linear combination of orientation and position co-ordinates, while
$p_{2k}$ is purely orientational. Dynamical evolution is therefore
completely determined by this Hamiltonian, preserving the symplectic
form $dp_{1}\wedge du_{1}+dp_{2}\wedge du_{2}$, and yielding a conserved
quantity $r_{k}$ known as a Casimir function \citep{PhysRevLett.71.3043,PhysRevLett.81.2399,olver2000applications,marsden2013introduction}.
The reduction of the state-space to a direct product of a symplectic
manifold and Casimir functions is known as Poisson dynamics, generalizing
Hamiltonian dynamics to odd-dimensional manifolds. The equations of
motion then take the form 
\[
\dot{q}_{ak}=\left\{ q_{ak},H_{k}\right\} ,\,\dot{p}_{ak}=\left\{ p_{ak},H_{k}\right\} 
\]
where $a\in\left\{ 1,2\right\} $ and the Poisson bracket is given
by
\[
\left\{ A_{k},H_{k}\right\} =\frac{\partial A_{K}}{\partial q_{ak}}\frac{\partial H_{k}}{\partial p_{ak}}-\frac{\partial H_{K}}{\partial q_{ak}}\frac{\partial A_{k}}{\partial p_{ak}}.
\]
 We now use this Hamiltonian to elucidate the stability behavior of
the Poisson orbits. The potential function 
\begin{align}
U_{k} & =\frac{1}{2}\left(\omega_{1}^{2}u_{1k}^{2}+\omega_{2}^{2}u_{2k}^{2}\right)\label{eq:potential}
\end{align}
may either be a saddle or parabolic, depending on the signs of $\omega_{1}^{2}$
and $\omega_{2}^{2}$. For $v>0$ we already deduced that $\omega_{1}^{2}<0$
while $\omega_{2}^{2}>0$ characterizing a saddle with unstable direction
aligned along $u_{1k}$. For $0>v>-s^{2}/2c$ both $\omega_{1}^{2},\omega_{2}^{2}<0$
resulting in an unstable parabolic potential. If $v$ is decreased further,
$\omega_{1}>0$ defining a saddle with unstable direction aligned
along $u_{2k}$. Contour plots of $U_{k}$ as $v$ is varied are presented
in Fig.(\ref{fig:Hkpic}a). Fig.(\ref{fig:Hkpic}b) depicts a plot of the Hessian function which shows that the potential only becomes parabolic in the range $0>v>-s^{2}/2c$. 

While the sedimenting lattice is always unstable for sedimenting active
spheres, we may consider an anisotropic body, or an active colloid
where the activity depends on spatial direction with longitudinal
and transverse components $v_{1},v_{2}$, yielding a modified dispersion
relation
\begin{equation}
\omega_{1}^{2}=s^{2}+2cv_{1},\quad\omega_{2}^{2}=-cv_{2}.\label{eq:potential-anisotropic}
\end{equation}
Now the potential may become positive-definite if $v_{1},v_{2}$ contain
opposite signatures, resulting in the appearance of stable bound orbits
with frequencies $\omega_{1},\omega_{2}$ in the longitudinal and
transverse directions respectively (see Fig.(\ref{fig:Hkpic}c-d)).
These frequencies are almost certainly non-commensurate and hence
indicate quasi-periodicity of the orbits. Such a situation may be
realized experimentally by considering a one-dimensional chain of
sedimenting triaxial bodies with minor axis aligned longitudinally
and major axis aligned transverse to the chain, or through imposition
of external fields in the transverse and longitudinal directions.
In the next section, we develop an equivalent continuum Cosserat theory
that models an active filament and compare this to the lattice-based
theory. 

\subsection{Continuum approximation}

\begin{figure}
\includegraphics[width=1\columnwidth]{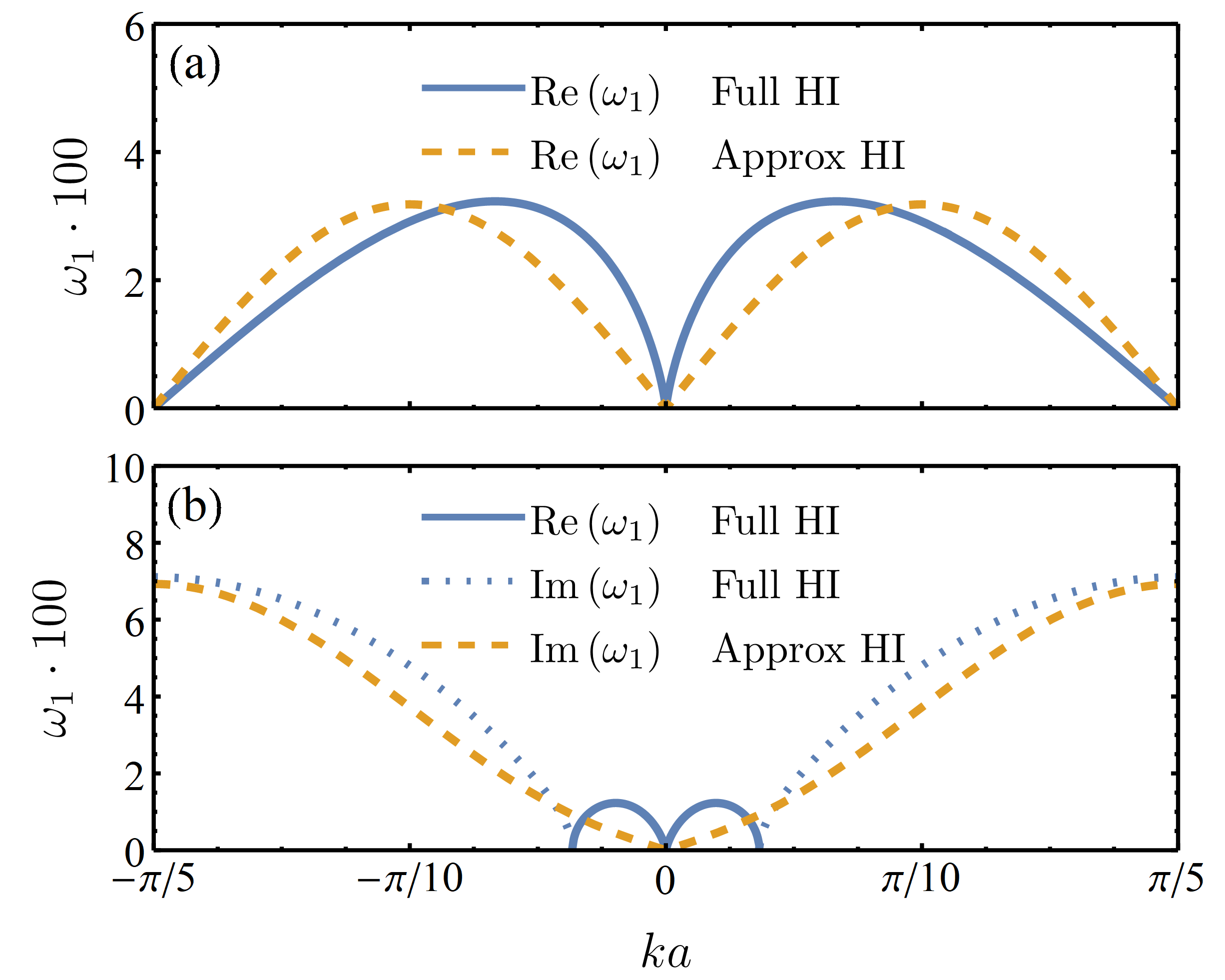}

\caption{\label{fig:1dcontinuumlinear} Comparison of instabilities in one-dimension
through plots of $\omega_{1}$. (a) Here, $v=0$ corresponding to
passive sedimenting spheres with entirely real and positive dispersion
relations. When long-ranged hydrodynamics are taken into account,
the dispersion relation forms a non-analytic cusp around $k=0$. For
$k>0$ the two dispersion relations qualitatively predict a similar
lattice instability (b) Here, $v=-1/20$ and the dispersion relations
no longer qualitatively agree at finite wavelength. The inclusion
of long-ranged hydrodynamics leads to a discrepancy between the approximate
dispersion relation, which is purely imaginary, and the full dispersion
relation which contains real and imaginary contributions. Therefore,
erroneous conclusions about the lattice stability may be drawn from
the bare continuum field theory, where lattice parameters are not
allowed to run with lengthscale.}
\end{figure}

We now derive approximate equations of motion for the dynamics of
an active Cosserat filament, the continuum analogue of the active
Cosserat chain. Here, displacement and orientation variables are now
functions of an additional continuous variable $x$ parameterizing
the filament in addition to time. As the lattice length scale $a\rightarrow0$,
we utilize the long-wavelength form of the dispersion relations to
construct our continuum theory. However, taking the long-wavelength
limit directly in Eq.(\ref{eq:Lfuncs}) leads to divergent lattice
sums, since both $s,c$ are non-analytic in their gradients at $k=0$ 
due to the long-ranged hydrodynamic forces present. We therefore truncate
these sums to $n=1$, corresponding to considering hydrodynamic effects
from nearest neighbors alone, while discarding long-ranged effects.
This yields analytic approximations

\begin{equation}
\begin{aligned}s\left(k\right)\equiv s & =i\frac{\lambda}{a}\sin ka\approx i\lambda k,\\
c\left(k\right)\equiv c & =\frac{\lambda b}{a^{2}}\left(\cos ka-1\right)\approx-\frac{1}{2}\lambda bk^{2}.
\end{aligned}
\label{eq:n-truncation}
\end{equation}
 The dispersion relations are then
\begin{equation}
\begin{alignedat}{1}\omega_{1}^{2} & =s^{2}+2cv\approx-\lambda\left(\lambda+vb\right)k^{2},\\
\omega_{2}^{2} & =-cv\approx\frac{1}{2}\lambda vbk^{2}.
\end{alignedat}
\label{eq:approx-lin-disp-1d}
\end{equation}
In analogy with Eq.(\ref{eq:explicit-1d-eqns}), the continuum equations
that produce this dispersion are given by

\begin{equation}
\begin{aligned}\dot{u}_{1}=- & \lambda\frac{\partial u_{3}}{\partial x}+vbq_{1},\enskip\dot{u}_{2}=vbq_{2},\enskip\dot{u}_{3}=\lambda\frac{\partial u_{1}}{\partial x},\\
\dot{q}_{1}=- & \lambda\frac{\partial^{2}u_{1}}{\partial x^{2}},\enskip\dot{q}_{2}=\frac{1}{2}\lambda\frac{\partial^{2}u_{2}}{\partial x^{2}}.
\end{aligned}
\label{eq:continuum-equations}
\end{equation}
These dynamics contain the conserved function
\[
\frac{\partial r}{\partial t}=\frac{\partial}{\partial t}\left(\lambda\frac{\partial^{2}u_{3}}{\partial x^{2}}+\lambda\frac{\partial q_{1}}{\partial x}\right)=0
\]
while the remaining variables close yielding the second order equations.
\begin{equation}
\frac{\partial^{2}u_{1}}{\partial t^{2}}=-\lambda\left(\lambda+vb\right)\frac{\partial^{2}u_{1}}{\partial x^{2}},\quad\frac{\partial^{2}u_{2}}{\partial t^{2}}=\frac{1}{2}\lambda vb\frac{\partial^{2}u_{2}}{\partial x^{2}}.\label{eq:continuum-2nd-order-1d}
\end{equation}
These equations also contain Poisson structure defined by the Hamiltonian
density

\begin{equation}
\begin{array}{cc}
\mathcal{H} & =\frac{1}{2}\left(\pi_{1}^{2}+\pi_{2}^{2}\right)+\frac{1}{2}\left[\alpha\left(\frac{\partial u_{1}}{\partial x}\right)^{2}+\beta\left(\frac{\partial u_{2}}{\partial x}\right)^{2}\right],\end{array}\label{eq:1d-hamiltonian-density}
\end{equation}
where 
\begin{equation}
\alpha=-\lambda\left(\lambda+vb\right),\quad\beta=\frac{1}{2}\lambda vb
\end{equation}
and now 
\begin{equation}
\pi_{1}=\partial_{t}u_{1},\,\pi_{2}=\partial_{t}u_{2}.\label{eq:piredef}
\end{equation}
Eqs.(\ref{eq:continuum-2nd-order-1d},\ref{eq:piredef}) together
are then equivalent to the set of first order equations given by
\begin{equation}
\dot{\boldsymbol{\pi}}\left(x\right)=\left\{ \boldsymbol{\pi}\left(x\right),H\right\} ,\quad\dot{\boldsymbol{u}}\left(x\right)=\left\{ \boldsymbol{u}\left(x\right),H\right\} ,\label{eq:1d-hamiltonian-deriv}
\end{equation}
where the Hamiltonian $H=\int\mathcal{H}\left(x\right)dx$ and we
have defined Poisson brackets
\[
\left\{ \mathcal{A}\left(x\right),H\right\} =\int\frac{\delta\mathcal{A}\left(x\right)}{\delta\boldsymbol{u}\left(x'\right)}\frac{\delta H}{\delta\boldsymbol{\pi}\left(x'\right)}-\frac{\delta H}{\delta\boldsymbol{u}\left(x'\right)}\frac{\delta\mathcal{A}\left(x\right)}{\delta\boldsymbol{\pi}\left(x'\right)}dx'.
\]
On the one hand, by performing a gradient expansion only keeping a finite
number of terms of the lattice sum, the resulting instability is of
a similar form to that studied by Crowley \citep{crowley1971viscosity}.
On the other hand, the dispersion relation derived by considering
the full hydrodynamics of an infinite sedimenting lattice is non-analytic,
resulting in the formation of a cusp as $k\rightarrow0$ (see Fig.(\ref{fig:1dcontinuumlinear})). 
As a consequence, the continuum theory is unable to reproduce
the behavior of the lattice-based model at zero wavenumber. Instabilities
of this type were first studied by Felderhof \citep{felderhof2003mesoscopic}. In addition,
for certain values of $v$ the dispersion relation derived from the
full lattice dynamics predicts a long-wavelength instability that
is absent from the dispersion relations of the approximate continuum
theory, as shown in Fig.(\ref{fig:1dcontinuumlinear}b). Unlike the
Felderhof instability, this discrepancy results from the truncation
of the analytic function performed in Eq.(\ref{eq:approx-lin-disp-1d})
and its effects may be included in the field theory by redefining
the lattice parameters to include a functional dependence on $k$
of the form
\begin{equation}
\begin{array}{cc}
\alpha\left(k\right)= & \omega_{1}^{2}\left(k\right)/k^{2},\\
\beta\left(k\right)= & \omega_{2}^{2}\left(k\right)/k^{2},
\end{array}
\end{equation}
where $\omega_{1},\omega_{2}$ depend on the bare lengthscales $\lambda,b$.
In this way, we see that it is possible to formulate a continuum theory
of the active Cosserat filament that reproduces the dispersion behavior
of the lattice-based model for $k\neq0$. This procedure is reminiscent
of a renormalization process whereby that the lattice parameters depend
on the lengthscale at which a system is probed. 

\section{Two dimensional lattice\label{sec:Sedimenting-active-2D}}

\begin{figure}
\includegraphics[width=1\columnwidth]{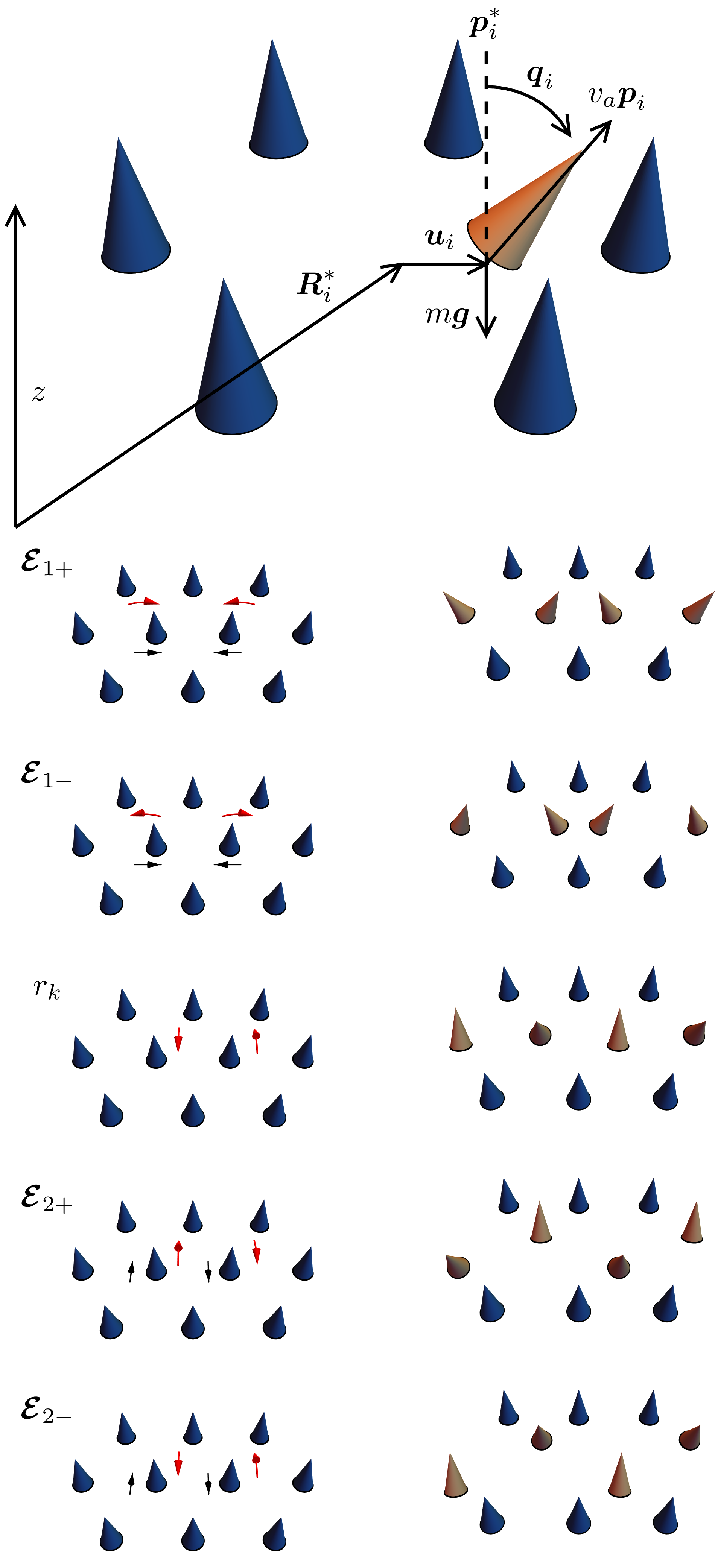}

\caption{\label{fig:Schematic-of-the-2d}The first panel shows a schematic
of the sedimenting two-dimensional crystal as specified by Eqs.(\ref{eq:linear-1st-order},\ref{eq:2d-lattice-specification}).
The remaining panels present a graphical depiction of the collective
modes evaluated at wavenumber $k=\pi/2$. The first three modes lie
in the space co-ordinatized by the triplet $\text{\ensuremath{\left(u_{\parallel k},q_{\parallel k},u_{3k}\right)}}$
while the remaining two modes lie in the space co-ordinatized by the
pair $\left(u_{\perp k},q_{\perp k}\right)$.}
\end{figure}
We now consider an infinite sedimenting two-dimensional lattice with
initial positions and orientations 

\begin{equation}
\boldsymbol{R}_{i}^{*}=\left(x_{i}^{*},y_{i}^{*},0\right),\quad\boldsymbol{p}_{i}^{*}=\left(0,0,1\right)\label{eq:2d-lattice-specification}
\end{equation}
where $x_{i}^{*},\,y_{i}^{*}$ define the initial lattice configuration.
The plane wave collective modes
\begin{align*}
\boldsymbol{u}_{i}\left(t\right) & =b\int\frac{d^{2}k}{4\pi^{2}}e^{i\left(k_{1}x_{i}^{*}+k_{2}y_{i}^{*}\right)}\boldsymbol{u}_{k}\left(t\right),\\
\boldsymbol{q}_{i}\left(t\right) & =\int\frac{d^{2}k}{4\pi^{2}}e^{i\left(k_{1}x_{i}^{*}+k_{2}y_{i}^{*}\right)}\boldsymbol{q}_{k}\left(t\right),
\end{align*}
where $\boldsymbol{k}=\left(k_{1},k_{2},0\right)$ may be inserted
into Eq.(\ref{eq:linear-evolution}) to yield lattice dynamics with
explicit form 

\begin{align}
\boldsymbol{J}_{k} & =\!\left(\begin{array}{ccc|ccc}
0 & 0 & -s\hat{k}_{1} & v & 0 & 0\\
0 & 0 & -s\hat{k}_{2} & 0 & v & 0\\
s\hat{k}_{1} & s\hat{k}_{2} & 0 & 0 & 0 & 0\\
\hline c\left(1-3\hat{k}_{1}\hat{k}_{1}\right) & -3c\hat{k}_{1}\hat{k}_{2} & 0 & 0 & 0 & 0\\
-3c\hat{k}_{1}\hat{k}_{2} & c\left(1-3\hat{k}_{2}\hat{k}_{2}\right) & 0 & 0 & 0 & 0\\
0 & 0 & 0 & 0 & 0 & 0
\end{array}\right)\label{eq:2dJk}
\end{align}
where we define $\hat{k}_{1}=k_{1}/k,\,\hat{k}_{2}=k_{2}/k$. The
lattice sums $s\left(\boldsymbol{k}\right)\equiv s,\,c\left(\boldsymbol{k}\right)\equiv c$
are defined in Appendix B. Once again, $s$ is purely imaginary
while $c$ is real and negative. The collective modes now consist
of positional polarizations $\tilde{\boldsymbol{u}}_{k},u_{3k}$ and
orientational polarizations $\boldsymbol{q}_{k}$ where
\begin{equation}
\tilde{\boldsymbol{u}}_{k}=\left(u_{1k},u_{2k},0\right)
\end{equation}
 are the components of $\boldsymbol{u}_{k}$ in the $xy$ plane, $u_{3k}$
is the component of $\boldsymbol{u}_{k}$ resolved parallel to gravity
and $\boldsymbol{q}_{k}=\tilde{\boldsymbol{q}}_{k}$ is a vector in
the $xy$ plane due to the constraint $\boldsymbol{p}^{*}\cdot\boldsymbol{q}_{k}=0$.
In this way, both $\tilde{\boldsymbol{u}}_{k},\tilde{\boldsymbol{q}}_{k}$
are vectors in the plane perpendicular to gravity and the final row
and column of $\boldsymbol{J}_{k}$ vanish leading to the presence
of a trivial null eigenvalue. The remaining dynamical degrees of freedom
yield the five non-trivial linear equations of motion
\begin{alignat}{1}
\begin{aligned}\dot{\tilde{\boldsymbol{u}}}_{k} & =-s\hat{\boldsymbol{k}}u_{3k}+v\boldsymbol{q}_{k},\quad\dot{u}_{3k}=s\hat{\boldsymbol{k}}\cdot\tilde{\boldsymbol{u}}_{k},\\
\dot{\boldsymbol{q}}_{k} & =c\left(\boldsymbol{I}-3\hat{\boldsymbol{k}}\hat{\boldsymbol{k}}\right)\cdot\tilde{\boldsymbol{u}}_{k}.
\end{aligned}
\label{eq:explicit-2s-eqns}
\end{alignat}
Again, non-symmetric coupling between positional and orientational
collective modes is indicative of the non-potential character of the
active forces and torques and this linear system decouples into two
subspaces. The first subspace consists of positional and orientational
modes longitudinal to $\hat{\boldsymbol{k}}$ given by 
\begin{equation}
	u_{\parallel k}=\hat{\boldsymbol{k}}\hat{\boldsymbol{k}}\cdot\boldsymbol{u}_{k},\quad
	q_{\parallel k}=\hat{\boldsymbol{k}}\hat{\boldsymbol{k}}\cdot\boldsymbol{q}_{k},
\end{equation}
respectively, in addition to the transverse mode $u_{3k}$ parallel
to gravity. These obey the linear equations
\begin{equation}
\frac{d}{dt}\begin{pmatrix}u_{\parallel k}\\
q_{\parallel k}\\
u_{3k}
\end{pmatrix}=\begin{pmatrix}0 & v & -s\\
-2c & 0 & 0\\
s & 0 & 0
\end{pmatrix}\begin{pmatrix}u_{1k}\\
q_{1k}\\
u_{3k}
\end{pmatrix}.
\end{equation}
While this equation looks identical to Eq.(\ref{eq:3dlindisp}), the
values for $s,c$ are different from those in Sec.(\ref{sec:Sedimenting-active-1D})
as the lattice sum is now over a two-dimensional crystal. However,
as this form is similar to that in  Sec.(\ref{sec:Sedimenting-active-1D}) we may repeat our analysis, deducing a conserved quantity orthogonal to the dynamical
flow given by
\begin{equation}
\mathcal{E}_{0}=2cu_{3k}+sq_{\parallel k}\equiv r_{k},\quad\dot{r}_{k}=0.
\end{equation}
This non-trivial conserved quantity again reflects the fact that local
activity can work to oppose passive clumping of the lattice \citep{crowley1976clumping}.
The dimensionality of the first subspace is therefore reduced by one
and is hence spanned by a pair of eigenmodes
\begin{equation}
\mathcal{E}_{\parallel\pm}=\pm\sqrt{-s^{2}-2cv}u_{\parallel k}+vq_{\parallel k}-su_{3k}.
\end{equation}
The positive sign denotes an in-phase position and orientation wave
directed along $\hat{\boldsymbol{k}}$ while the negative sign denotes
its anti-phase counterpart. These eigenmodes obey the harmonic equation
\begin{equation}
\ddot{\mathcal{E}}_{\parallel\pm}=-\omega_{\parallel}^{2}\mathcal{E}_{\parallel\pm},\,\omega_{\parallel}^{2}=s^{2}+2cv.
\end{equation}
Again, we see the appearance of stable harmonic waves of frequency
$\omega_{\parallel}$ when $v<-s^{2}/2c$, which is negative, such
that $\omega_{\parallel}^{2}>0$. Conversely, when $v>-s^{2}/2c$
the square of the frequency $\omega_{\parallel}^{2}<0$ implies
the presence of an exponentially growing mode $\mathcal{E}_{\parallel+}$
while the conjugate mode $\mathcal{E}_{\parallel-}$ is exponentially
decaying. The solution for passive spheres is obtained by taking $v=0$
which immediately implies that passive sedimenting two-dimensional
lattices are unstable \citep{crowley1976clumping}. 

The second subspace consists of positional and orientational modes
transverse to both $\hat{\boldsymbol{k}}$ and $u_{3k}$ consisting
of the linear perturbation modes $u_{\perp k}=\left(\boldsymbol{I}-\hat{\boldsymbol{k}}\hat{\boldsymbol{k}}\right)\cdot\boldsymbol{u}_{k},q_{\perp k}=\left(\boldsymbol{I}-\hat{\boldsymbol{k}}\hat{\boldsymbol{k}}\right)\cdot\boldsymbol{q}_{k}$
which obey the equations
\begin{equation}
\frac{d}{dt}\begin{pmatrix}u_{\perp k}\\
q_{\perp k}
\end{pmatrix}=\begin{pmatrix}0 & v\\
c & 0
\end{pmatrix}\begin{pmatrix}u_{\perp k}\\
q_{\perp k}
\end{pmatrix}.
\end{equation}
This leads to the appearance of two linearly independent transverse
position-orientation coupled eigenmodes given by
\begin{equation}
\mathcal{E}_{\perp\pm}=\pm\sqrt{cv}u_{\perp k}+vq_{\perp k}.
\end{equation}
Again, the positive sign denotes a transverse in-phase position and
orientation wave while the negative sign denotes its anti-phase counterpart.
These eigenmodes obey a different harmonic equation given by
\begin{equation}
\ddot{\mathcal{E}}_{\perp\pm}=-\omega_{\perp}^{2}\mathcal{E}_{\perp\pm},\,\omega_{\perp}^{2}=-cv.
\end{equation}
In this case, it is clear that for $v>0$ we obtain wavelike solutions
of frequency $\omega_{\perp}$. However, for $v<0$ we see that the
square of the frequency $\omega_{\perp}^{2}<0$ implying the presence
of an exponentially growing mode $\mathcal{E}_{\perp+}$ while the
conjugate mode $\mathcal{E}_{\perp-}$ is exponentially decaying.
For passive spheres, $\omega_{\perp}=0$ and transverse perturbations
are neither stable nor unstable. Together, we see that there is no
value of $v$ for which both longitudinal and transverse modes are
stable, and thus an exponentially growing solution will always be
present. Furthermore, for the range of parameter values $0>v>-s^{2}/2c$
both transverse and longitudinal modes $\mathcal{E}_{\parallel+},\mathcal{E}_{\perp+}$
are unstable. Where previously the one-dimensional chain explicitly
broke rotational symmetry, here the discrete rotational symmetry of
the two-dimensional crystal is only broken upon the application of
a wave-like perturbation with wavemode $\boldsymbol{k}$. In this
way, the longitudinal and transverse directions to this wavevector
provide a natural co-ordinate system, diagonalizing Eq.(\ref{eq:2dJk}) and
rendering it in the same form as Eq.(\ref{eq:1dJk}). In the next
section, we show that the two-dimensional lattice also admits symplectic
structure allowing us to reformulate the stability criteria through
the construction of a scalar Hamiltonian function as done previously. 

\subsection{Poisson structure}

We start by making the co-ordinate transformations
\begin{align}
p_{\parallel k}=-su_{3k}+vq_{\parallel k},\quad p_{\perp k}=vq_{\perp k}
\end{align}
which immediately yields the Poisson structure
\begin{equation}
\frac{d}{dt}\begin{pmatrix}u_{\parallel k}\\
u_{\perp k}\\
p_{\parallel k}\\
p_{\perp k}\\
r_{k}
\end{pmatrix}=\left(\begin{array}{cccc|c}
0 & 0 & 1 & 0 & 0\\
0 & 0 & 0 & 1 & 0\\
-1 & 0 & 0 & 0 & 0\\
0 & -1 & 0 & 0 & 0\\
\hline 0 & 0 & 0 & 0 & 0
\end{array}\right)\boldsymbol{\nabla}H_{k},\label{eq:poisson-structure-2d}
\end{equation}

\[
\boldsymbol{\nabla}=\begin{pmatrix}\partial/\partial u_{\parallel k}, & \partial/\partial u_{\perp k}, & \partial/\partial p_{\parallel k}, & \partial/\partial p_{\perp k}, & \partial/\partial r_{k}\end{pmatrix}
\]
with Hamiltonian
\begin{equation}
\begin{array}{cc}
H_{k} & =\frac{1}{2}\left(p_{\parallel k}^{2}+p_{\perp k}^{2}\right)+\frac{1}{2}\left(\omega_{1}^{2}u_{\parallel k}^{2}+\omega_{2}^{2}u_{\perp k}^{2}\right),\end{array}
\end{equation}
where 
\begin{equation}
\omega_{1}^{2}=\left(s^{2}+2cv\right)\quad\omega_{2}^{2}=-cv.
\end{equation}
Here, $p_{\parallel k},\,p_{\perp k}$ are momentum-like variables
conjugate to $u_{\parallel k},\,u_{\perp k}$. $p_{\parallel k}$
consists of a linear combination of orientation and position co-ordinates,
while $p_{\perp k}$ is purely orientational. Dynamical evolution
is therefore completely determined by this Hamiltonian, preserving
the symplectic form $dp_{\parallel}\wedge du_{\parallel}+dp_{\perp}\wedge du_{\perp}$,
and yielding a conserved Casimir function $r_{k}$ \citep{PhysRevLett.71.3043,PhysRevLett.81.2399,olver2000applications,marsden2013introduction}.
The equations of motion may then be written in the form 
\[
\dot{q}_{ak}=\left\{ q_{ak},H_{k}\right\} ,\,\dot{p}_{ak}=\left\{ p_{ak},H_{k}\right\} 
\]
where $a\in\left\{ \parallel,\perp\right\} $ and the Poisson bracket
is given by
\[
\left\{ A_{k},H_{k}\right\} =\frac{\partial A_{k}}{\partial q_{ak}}\frac{\partial H_{k}}{\partial p_{ak}}-\frac{\partial H_{k}}{\partial q_{ak}}\frac{\partial A_{k}}{\partial p_{ak}}.
\]
 We now use this Hamiltonian to elucidate the stability behavior of
the Poisson orbits. Once again, the potential function 
\begin{align}
U_{k} & =\frac{1}{2}\left(\omega_{\parallel}^{2}u_{\parallel k}^{2}+\omega_{\perp}^{2}u_{\perp k}^{2}\right)\label{eq:potential-1}
\end{align}
is parabolic when $0>v>-s^{2}/2c$ such that both $\omega_{\parallel}^{2},\omega_{\perp}^{2}<0$
and a saddle-point for all other values of $v$, whereby either $\omega_{\parallel}^{2}<0$ and $\omega_{\perp}^{2}>0$ or $\omega_{\parallel}^{2}>0$ and $\omega_{\perp}^{2}<0$.
 We therefore see that, for
isotropic activity, the lattice is unstable for all values of $v$.
However, if the crystal is composed of particles that deform anisotropically or are polarized upon macroscopic perturbations of 
the crystal, the activity may decompose into unique parts
$v_{\parallel},v_{\perp}$ transverse and perpendicular to the applied
perturbation. This yields a modified dispersion relation
\begin{equation}
\omega_{\parallel}^{2}=s^{2}+2cv_{\parallel},\quad\omega_{\perp}^{2}=-cv_{\perp}.\label{eq:potential-anisotropic-1}
\end{equation}
In this way, the potential $U_{k}$ may become positive-definite for
certain values of $v_{\parallel}$ and $v_{\perp}$ resulting in stable orbits.

\subsection{Continuum approximation}

We now derive approximate continuum equations of motion for the dynamics
of the active Cosserat sheet, the continuum analogue of the two-dimensional
active Cosserat crystal. Here, displacement and orientation variables
are now functions of the two continuous variables $x,y$ in addition
to time. Following Sec.(\ref{sec:Sedimenting-active-1D}b), the hydrodynamics are truncated
to only include nearest neighbor interactions yielding
\begin{equation}
\begin{aligned}\boldsymbol{s}\left(\boldsymbol{k}\right) & =i\frac{\lambda}{a}\left(\hat{\boldsymbol{x}}\sin k_{1}a+\hat{\boldsymbol{y}}\sin k_{2}a\right)\approx i\lambda\boldsymbol{k},\\
c\left(\boldsymbol{k}\right) & =\frac{\lambda b}{a^{2}}\left[\cos k_{1}a+\cos k_{2}a-2\right]\approx-\frac{1}{2}\lambda bk^{2}.
\end{aligned}
\end{equation}
 $\boldsymbol{s}\left(\boldsymbol{k}\right)$ is a vectoral quantity
which is proportional to the applied perturbation direction $\boldsymbol{k}$
when the corresponding sum is taken over all lattice points (see Appendix
B). When this sum is truncated to nearest neighbors, the underlying
point group is exposed by the lattice vectors $\hat{\boldsymbol{x}},\hat{\boldsymbol{y}}$.
However, in the long-wavelength limit the crystal appears isotropic
and $\boldsymbol{s}$ once again can only depend on $\boldsymbol{k}$. Using Eq.(\ref{eq:explicit-2s-eqns}) immediately leads to the continuum
equations 
\begin{equation}
\begin{aligned}\dot{\tilde{\boldsymbol{u}}} & =-\lambda\boldsymbol{\nabla}u_{z}+vb\boldsymbol{q},\enskip\dot{u}_{3}=\lambda\boldsymbol{\nabla}\cdot\tilde{\boldsymbol{u}},\\
\dot{\boldsymbol{q}} & =\frac{1}{2}\lambda\left(\nabla^{2}\tilde{\boldsymbol{u}}-3\boldsymbol{\nabla}\boldsymbol{\nabla}\cdot\tilde{\boldsymbol{u}}\right),
\end{aligned}
\end{equation}
where $\boldsymbol{\nabla}=\left(\partial_{x},\partial_{y},0\right)$
denotes a gradient operator taken in the plane perpendicular to gravity.
These equations contain a conserved function
\begin{equation}
\begin{aligned}\frac{\partial r}{\partial t}=\frac{\partial}{\partial t}\left(\nabla^{2}u_{3}+\boldsymbol{\nabla}\cdot\boldsymbol{q}_{k}\right)=0\end{aligned}
,
\end{equation}
and can be closed yielding the second order evolution
\begin{equation}
\ddot{\tilde{\boldsymbol{u}}}=-\lambda\left(\lambda+\frac{3}{2}vb\right)\boldsymbol{\nabla}\boldsymbol{\nabla}\cdot\tilde{\boldsymbol{u}}+\frac{1}{2}\lambda vb\nabla^{2}\tilde{\boldsymbol{u}}.\label{eq:2duddot}
\end{equation}
This equation is identical to that of a linear elastic medium where
$-\lambda\left(\lambda+3vb/2\right),\lambda vb/2$ are the Lamé parameters.
However, unlike an elastic medium, the compression modulus of the
active medium may be negative, resulting in further contraction and
instability upon application of external pressure or shearing. We
can define a Hamiltonian density
\begin{equation}
\begin{array}{cc}
\mathcal{H}= & \frac{1}{2}\boldsymbol{\pi}^{2}+\frac{1}{2}\left[\alpha\left(\boldsymbol{\nabla}\cdot\tilde{\boldsymbol{u}}\right)^{2}+\beta\left(\boldsymbol{\nabla}\tilde{\boldsymbol{u}}\right)^{2}\right],\end{array}\label{eq:2d-hamiltonian-density}
\end{equation}
where
\begin{equation}
\alpha=-\lambda\left(\lambda+\frac{3}{2}vb\right),\quad\beta=\frac{1}{2}\lambda vb
\end{equation}
and now 
\begin{equation}
\boldsymbol{\pi}=\partial_{t}\boldsymbol{u}.\label{eq:pivecdt}
\end{equation}
Eqs.(\ref{eq:2duddot},\ref{eq:pivecdt}) together are then equivalent
to the set of first order equations given by
\begin{equation}
\dot{\boldsymbol{\pi}}\left(\boldsymbol{x}\right)=\left\{ H,\boldsymbol{\pi}\left(\boldsymbol{x}\right)\right\} ,\quad\dot{\boldsymbol{u}}\left(\boldsymbol{x}\right)=\left\{ H,\boldsymbol{u}\left(\boldsymbol{x}\right)\right\} .\label{eq:1d-hamiltonian-deriv-1}
\end{equation}
where the Hamiltonian $H=\int\mathcal{H}\left(\boldsymbol{x}\right)d^{2}\boldsymbol{x}$
and we have defined Poisson brackets given by
\[
\left\{ H,\mathcal{A}\left(\boldsymbol{x}\right)\right\} =\int\frac{\delta H}{\delta\boldsymbol{u}\left(\boldsymbol{x}'\right)}\frac{\delta\mathcal{A}\left(\boldsymbol{x}\right)}{\delta\boldsymbol{\pi}\left(\boldsymbol{x}'\right)}-\frac{\delta H}{\delta\boldsymbol{\pi}\left(\boldsymbol{x}'\right)}\frac{\delta\mathcal{A}\left(\boldsymbol{x}\right)}{\delta\boldsymbol{u}\left(\boldsymbol{x}'\right)}d^{2}\boldsymbol{x}'.
\]
The Hamiltonian density is manifestly rotationally
invariant with geometric factors $\alpha,\beta$ dependent on the
underlying microstructure. As shown previously, stable sedimentation
may only occur if activity or microscopic deformation is induced
by the applied perturbation. This is similar to many microstructural
materials that display microscopic anisotropy upon application of
macroscopic strain or shearing. Once again, this can result in two
anisotropy parameters, $v_{1},v_{2}$, along and transverse to the
applied deformation which, through similar analysis as done previously,
may yield stable Poisson orbits. A possible system that may exhibit
this behavior is sedimenting arrays of electrically charged fluid
droplets which deform under lattice compression. The dispersion relations
formulated this way display a similar discrepancy compared to the
full lattice-based dynamics as discussed in Sec.(\ref{sec:Sedimenting-active-1D}b).
This results in both the running of the lattice parameters $\alpha,\beta$
with lengthscale, which can be included in the continuum theory, and
the appearance of a Felderhof instability \citep{felderhof2003mesoscopic}
which cannot be included in the continuum theory.

\section{Discussion}

On the one hand, the appearance of Poisson structure in the active
Cosserat crystal may be expected, since it can be shown that any odd-dimensional
linear system yields a Poisson structure \citep{estabrook1975geometric}.
On the other hand, many previous examples of non-linear symplectic
structure arising in the field of active matter share a common feature,
namely the identification of orientation variables as conjugate momenta
to displacements \citep{hocking_1964,PhysRevLett.108.218104,zottl2013periodic,lushi2015periodic,stark2016swimming,Shelleyetal,tallapragada2019chaotic,chajwa2019kepler,PhysRevLett.124.088003}.
We suspect a similar structure is present in previous works involving
two-body problems and that additional Lie symmetries could be found,
reducing the dynamical space to a single conjugate pair. This implies
a certain non-trivial ubiquity of Poissonian dynamics within the field
of active matter, even at non-linear order, which appears to be a
rich area for further research. The recurrence of symplectic structure
can be rationalized through the following heuristic argument: many
active system share a common motif with regards to time-evolution
whereby position variables are updated according
to orientational state while orientation variables are
updated according to positional state. The absence of an on-diagonal
response, as seen in Eqs.(\ref{eq:1dJk},\ref{eq:2dJk}), implies
conservation of phase volume which, in two dimensions, is
a conserved symplectic form. Indeed, both the one- and two-dimensional
Cosserat crystals can be decoupled into two- and three- dimensional
Hamiltonian and Poissonian subsystems of the entire state-space, comprised
of transverse and longitudinal dynamics respectively, whereby the
conjugate momenta are monotonically dependent on orientation variables.

Furthermore, the presence of a non-analytic point at zero wavenumber
in both the one- and two-dimensional active Cosserat crystals is indicative
of an instability of the type studied by Felderhof \citep{felderhof2003mesoscopic}
and results due to the presence of long-ranged hydrodynamic forces.
This feature cannot be modeled well by gradient expansion methods,
which amount to a nearest-neighbor approximation of the full theory
\citep{crowley1971viscosity,crowley1976clumping,lahiri1997steadily},
and leads to a discrepancy between continuum theories and discrete lattice-based
analysis at wavenumber $k=0$. 

To conclude this section, we remark that this Poissonian interpretation
of the sedimentation behavior of active Cosserat crystals can be used
to understand hydrodynamically-mediated crystallization at fluid boundaries
\citep{singh2016crystallization,maass2016swimming,thutupalli2018FIPS,caciagli2020controlled}.
In analogy with our previous work, \citep{PhysRevLett.124.088003},
we expect the boundary to provide a type of forcing through hydrodynamic
interactions, which may be damped by the inclusion of further external
forces and torques. We defer investigation into the possible steady-state
behavior to future work \citep{bolithoCossCrystal}.

\section{Conclusion}

We investigated the sedimentation behavior of both the sedimenting
one- and two-dimensional active Cosserat crystal in an overdamped
external medium and, using a virtual power principle \citep{germain1973method},
derive the geometrical equations of motion for the lattice of active
uniaxial colloidal particles. In the presence of an overdamped external
medium, inertial forces are negligible and the lattice dynamics evolve
on a five-dimensional state-space of translation and orientation modes.
Remarkably, this state-space in endowed with Poisson structure \citep{PhysRevLett.71.3043,PhysRevLett.81.2399,olver2000applications,marsden2013introduction}
even in the absence of inertial and conservative forces, with coupled
sedimentation-orientation co-ordinates playing the role of conjugate
momenta to the lattice displacements. We identify conserved Casimir
functions and a simple harmonic Hamiltonian with activity dependent
frequencies. Together, these form a symplectic foliation of the dynamical
state-space, thereby reducing the problem to one of Hamiltonian dynamics.
This is exploited to reveal the presence of stable position-orientation-sedimentation
coupled limit-cycle behavior, which is shown to occur in both one-
and two-dimensions, in the presence of anisotropic activity. 
\begin{acknowledgments}
We acknowledge the EPSRC (AB) and the Isaac Newton Trust (RA) for
support. We thank Prof. M. E. Cates and Prof. R. E. Goldstein for
helpful discussions and critical remarks. We thank Prof. R. E. Goldstein
for bringing \citep{solovev2021lagrangian} to our attention.
\end{acknowledgments}

\bibliographystyle{apsrev4-2}

\widetext
\selectlanguage{english}%

\appendix

\section{Lattice sums in one dimension\label{sec:1D-lattice-sums}}

The Greens function of an unbounded viscous medium $\boldsymbol{G}$
given by 
\begin{equation}
\boldsymbol{G}\left(\boldsymbol{R}_{i},\boldsymbol{R}_{j}\right)=\frac{1}{8\pi\eta r}\left(\boldsymbol{I}+\hat{\boldsymbol{r}}_{ij}\hat{\boldsymbol{r}}_{ij}\right)\label{eq:unbound-greens}
\end{equation}
where $\boldsymbol{r}_{ij}=\boldsymbol{R}_{i}-\boldsymbol{R}_{j}$and
$\hat{\boldsymbol{r}}_{ij}=\boldsymbol{r}_{ij}/r_{ij}$ which may
also be written as 
\begin{equation}
\boldsymbol{G}\left(\boldsymbol{R}_{i},\boldsymbol{R}_{j}\right)=\left(\nabla^{2}\boldsymbol{I}-\boldsymbol{\nabla}\boldsymbol{\nabla}\right)r_{ij}\label{eq:simpleG}
\end{equation}
The lattice sums required to calculate the hydrodynamical interactions
of the one-dimensional chain can be evaluated using
\begin{align}
\sum_{l}\left[\frac{\partial\boldsymbol{\mu}_{il}^{TT}}{\partial\boldsymbol{R}_{j}}\cdot m\boldsymbol{g}\right]_{*} & =\sum_{l}\left[\frac{\partial}{\partial\boldsymbol{R}_{j}}\frac{1}{8\pi\eta r_{il}}\left(m\boldsymbol{g}+\hat{\boldsymbol{r}}_{il}\left(\hat{\boldsymbol{r}}_{il}\cdot m\boldsymbol{g}\right)\right)\right]_{*}\nonumber \\
 & =-\frac{mg}{8\pi\eta r_{ij}^{2}}\left[-\left(\hat{\boldsymbol{z}}+\hat{\boldsymbol{r}}_{ij}\left(\hat{\boldsymbol{r}}_{ij}\cdot\hat{\boldsymbol{z}}\right)\right)\hat{\boldsymbol{r}}_{ij}+\boldsymbol{I}\left(\hat{\boldsymbol{r}}_{ij}\cdot\hat{\boldsymbol{z}}\right)+\hat{\boldsymbol{r}}_{ij}\hat{\boldsymbol{z}}\right]_{*}\nonumber \\
 & =\frac{mg}{8\pi\eta r_{ij}^{2}}\left[\hat{\boldsymbol{z}}\hat{\boldsymbol{r}}_{ij}-\hat{\boldsymbol{r}}_{ij}\hat{\boldsymbol{z}}\right]
\end{align}
where, by parity, $\sum_{l}\hat{\boldsymbol{r}}_{jl}=0$. Using a
similar procedure and setting $\boldsymbol{r}_{ij}^{*}=\boldsymbol{R}_{i}^{*}-\boldsymbol{R}_{j}^{*}=na\hat{\boldsymbol{x}}$,
the following lattice sums can then be evaluated:

\begin{align}
\sum_{ln}e^{ikna}\left[\frac{\partial\boldsymbol{\mu}_{il}^{TT}}{\partial\boldsymbol{R}_{j}}\cdot m\boldsymbol{g}\right]_{*} & =\frac{img}{4\pi\eta a^{2}}\left(\hat{\boldsymbol{z}}\hat{\boldsymbol{x}}-\hat{\boldsymbol{x}}\hat{\boldsymbol{z}}\right)\sum_{n=1}^{\infty}\frac{\sin nka}{n^{2}}\\
\sum_{ln}e^{ikna}\left[\hat{\boldsymbol{z}}\times\frac{\partial\boldsymbol{\mu}_{il}^{RT}}{\partial\boldsymbol{R}_{j}}\cdot m\boldsymbol{g}\right]_{*} & =\frac{mg}{4\pi\eta a^{3}}\left(\boldsymbol{I}-\hat{\boldsymbol{z}}\hat{\boldsymbol{z}}-3\hat{\boldsymbol{x}}\hat{\boldsymbol{x}}\right)\sum_{n=1}^{\infty}\left(\frac{\cos nka}{n^{3}}-\frac{1}{n^{3}}\right)
\end{align}

\section{Lattice sums in two dimensions\label{sec:2-dimensional-Fourier-transform}}

Taking the lattice to be rectangular with $\boldsymbol{r}_{ij}=na\hat{\boldsymbol{x}}+ma\hat{\boldsymbol{y}}$,
the two-dimensional lattice sums required are then
\begin{align}
\sum_{lnm}e^{i\left(k_{1}n+k_{2}m\right)a}\left[\frac{\partial\boldsymbol{\mu}_{il}^{TT}}{\partial\boldsymbol{R}_{j}}\cdot m\boldsymbol{g}\right]_{*} & =\frac{img}{4\pi\eta a^{2}}\sum_{nm}^{\infty}\left[\hat{\boldsymbol{z}}\hat{\boldsymbol{r}}_{ij}-\hat{\boldsymbol{r}}_{ij}\hat{\boldsymbol{z}}\right]_{*}\frac{\sin\left(nk_{1}+mk_{2}\right)a}{n^{2}+m^{2}}=-\frac{ia}{\lambda}\left(\hat{\boldsymbol{z}}\boldsymbol{s}\left(\boldsymbol{k}\right)-\boldsymbol{s}\left(\boldsymbol{k}\right)\hat{\boldsymbol{z}}\right),\\
\sum_{lnm}e^{i\left(k_{1}n+k_{2}m\right)a}\left[\hat{\boldsymbol{z}}\times\frac{\partial\boldsymbol{\mu}_{il}^{RT}}{\partial\boldsymbol{R}_{j}}\cdot m\boldsymbol{g}\right]_{*} & =\frac{mg}{4\pi\eta a^{3}}\sum_{nm}^{\infty}\left[\boldsymbol{I}-\hat{\boldsymbol{z}}\hat{\boldsymbol{z}}-3\hat{\boldsymbol{r}}_{ij}\hat{\boldsymbol{r}}_{ij}\right]_{*}\frac{\cos\left(nk_{1}+mk_{2}\right)a-1}{\left(n^{2}+m^{2}\right)^{3/2}}=\frac{a^{2}}{\lambda b}c\left(\boldsymbol{k}\right)\left(\boldsymbol{I}-\hat{\boldsymbol{z}}\hat{\boldsymbol{z}}-3\hat{\boldsymbol{k}}\hat{\boldsymbol{k}}\right),
\end{align}
where $\sum_{nm}^{\infty}$ implies summation over all lattice points
$n>0,-\infty<m<\infty$ in the half-plane $\Gamma$ and $\hat{\boldsymbol{r}}_{ij}=\left(n,m\right)/\sqrt{n^{2}+m^{2}}$.
These sums are conditionally convergent \citep{campbell1963existence,borwein2013lattice}
yielding finite expressions for $\boldsymbol{s}\left(\boldsymbol{k}\right),c\left(\boldsymbol{k}\right)$.
In particular, the sum
\begin{equation}
\boldsymbol{s}\left(\boldsymbol{k}\right)=\sum_{nm}^{\infty}\hat{\boldsymbol{r}}_{ij}\frac{\sin\left(nk_{1}+mk_{2}\right)a}{n^{2}+m^{2}}=s\left(\boldsymbol{k}\right)\hat{\boldsymbol{k}}
\end{equation}
since $\sin\left(\boldsymbol{k}\cdot\hat{\boldsymbol{r}}_{ij}\right)=0$
when $\hat{\boldsymbol{r}}_{ij}\perp\boldsymbol{k}$ and therefore
only components parallel to the wavevector contribute when the sum
is taken over all lattice points. 

\selectlanguage{american}%

\end{document}